\documentclass[%
aip,
amsmath,amssymb,
reprint,%
]{revtex4-1}
\usepackage{graphicx} 
\usepackage{subfigure}
\usepackage{multirow}
\usepackage{fancyhdr}
\usepackage{longtable}
\usepackage{parskip}
\usepackage[T1]{fontenc}
\usepackage{dcolumn} 
\usepackage{bm}       
\usepackage{amsfonts}  
\usepackage{amsmath}   
\usepackage{amssymb}   
\usepackage[]{hyperref}

\usepackage[utf8]{inputenc}
\usepackage[T1]{fontenc}
\usepackage{mathptmx}
\usepackage{etoolbox}
\usepackage{xcolor}

\begin{document}
	
\title{Synchronization of two coupled massive  oscillators in the time-delayed Kuramoto model}

\author{Esmaeil Mahdavi}
\affiliation{ Department of Physics, Institute for Advanced Studies in Basic Sciences (IASBS), Zanjan 45137-66731, Iran}

\author{Mina Zarei}
\affiliation{ Department of Physics, Institute for Advanced Studies in Basic Sciences (IASBS), Zanjan 45137-66731, Iran}

\author{Farhad Shahbazi} 
\affiliation{ Department of Physics, Isfahan University of Technology, Isfahan 84156-83111, Iran}

\begin{abstract}  

We examine the impact of time delay on two coupled massive oscillators within the second-order Kuramoto model, which is relevant to the operations of real-world networks that rely on signal transmission speed constraints. Our analytical and numerical exploration shows that time delay can cause multi-stability within phase-locked solutions, and the stability of these solutions decreases as inertia increases. In addition to phase-locked solutions, we discovered non-phase-locked solutions that exhibit periodic and chaotic behaviors, depending on the amount of inertia and time delay. 

\end{abstract}

\maketitle
 
\begin{quotation}
Synchronization of two over-damped phase oscillators coupled by the time-delayed Kuramoto interaction has been studied, using linear stability analysis~\cite{ schuster1989mutual}. For the case of two identical oscillators, It has been shown, that for all values of time delay, there are one or several phase-locked solutions. Here,  we show that inertia tends to destabilize some stable solutions of the first-order Kuramoto model and convert them to solutions with periodic or chaotic dynamics. Moreover, introducing a difference in the natural frequencies of the two oscillators leads to a rich variety of dynamical behaviors. 
\end{quotation}

\section{Introduction}
 The tendency to synchronize may be the most mysterious and widespread phenomenon observed within the realms of nature, engineering, and social life~\cite{strogatz2008sync,nijmeijer2003synchronization,mirollo1990synchronization,tyson1973some,blasius1999complex,pikovsky2003synchronization}. This remarkable propensity profoundly shapes the dynamics and functioning of the systems consisting of interacting units. Extensive research has been dedicated to modeling synchronization, where mathematical and computational models are developed to describe the dynamics and mechanisms of synchronized behavior. These models consider a diverse range of factors, including coupling topology, frequency distribution, stochasticity and noise, adaptation, and time delay, among others~\cite{arenas2008synchronization,pecora1998master,nakao2007noise,cao2006adaptive, schuster1989mutual}. By incorporating these factors, researchers can create synchronization models that closely resemble the complex dynamics observed in real-world systems and phenomena.

 Time delay is a fundamental factor that significantly influences synchronization dynamics. It arises from the inherent limitations in signal transfer speed, resulting in a temporal lag between the transmission and reception of information or signals in a system. This limitation is present in various applications, such as neural networks, arrays of lasers, electronic circuits, and microwave oscillators~\cite{kerszberg1990synchronization,waibel2013phoneme,kozyreff2000global,reddy2000experimental,yeung1999time}. The effects of time delays on synchronization can indeed vary depending on the magnitude of the delays and the underlying dynamics of the system. Time delays can have both stabilizing and destabilizing effects on synchronization. They can enhance synchronization by allowing units to adapt and adjust their behavior based on delayed information from other units. This adaptability promotes coordination and coherence. However, excessive delays or specific delay configurations can have the opposite effect, leading to instability or even the loss of synchronization~\cite{yeung1999time,ziaeemehr2020frequency}. 
Time delay can lead to the emergence of multistability by altering the stability landscape of the system~\cite{madadi2018delay,schuster1989mutual,yeung1999time,choi2000synchronization}. Moreover, the amount of information transfer between two oscillating neural populations with a mismatch in their frequencies can be determined by their connection delay~\cite{pariz2021transmission}. Introducing an intermediate population and incorporating time delay has been shown to facilitate the synchronization of two unconnected neuronal populations~\cite{vicente2008dynamical}. In Small-World networks, the time delay can lead to enhanced synchronization, discontinuous phase transitions, and the emergence of chimera states~\cite{ameli2021time}.
 
The Kuramoto model is a well-known and widely used model for studying synchronization. It describes the phase evolution of coupled oscillators in a system using first-order nonlinear differential equations~\cite{kuramoto1975self,kuramoto2012chemical}. However, the original version of the Kuramoto model did not match the observations in some biological systems. Indeed, it predicted a rapid achievement of partial synchronization, contrary to the slower dynamics observed in synchronization of certain species of fireflies~\cite{ermentrout1991adaptive}. To address this discrepancy, Ermentrout introduced the frequency adaptation into the phase dynamics, resulting in a group of second-order coupled differential equations~\cite{ermentrout1991adaptive}. This modification retards the speed of reaching partial synchronization, aligning more closely with the observed dynamics. 
The second-order Kuramoto model incorporates inertia into the original Kuramoto model, resulting in a new model with valuable applications in fields such as power grids and Josephson junction arrays~\cite{filatrella2008analysis,rohden2012self,rohden2014impact,grzybowski2016synchronization,trees2005synchronization}. This introduces significant changes to the system dynamics, including alterations to the nature of the phase transition~\cite{tanaka1997first}. Despite this, the full implications of these changes remain relatively unexplored, largely due to limited research and investigation~\cite{dorfler2011critical, kachhvah2017multiplexing}. 

 Following Schuster and Wagner's linear stability analysis of two over-damped coupled phase oscillators with time-delayed
  Kuramoto interaction~\cite{ schuster1989mutual}, we explore the effect of time delay and inertia on the second-order Kuramoto model, incorporating two phase oscillators.
 Our investigation reveals that time delay induces multi-stability in the system. We demonstrate that increasing the inertia leads to the disruption of the phase-locked solutions. Furthermore, apart from the phase-locked solutions, we observe periodic and chaotic behaviors. 
 
 The paper is organized as follows: In the first section, we conduct a detailed mathematical analysis to examine the phase-locked solutions in the system and investigate their stability. Section~\ref{results} presents the numerical findings of our study. We discuss both the phase-locked and non-phase-locked solutions. We analyze their characteristics and discuss the implications of these different types of solutions. In the final section, we summarize our results.
 
\section{method}
\label{model}
Consider two-phase oscillators (rotors) with equal rotational inertia $M_1=M_2=M$ derived by the external torques $I_1$ and $I_2$,  and coupled by the time-delayed Kuramoto interaction.  Considering the damping, the dynamical equations of this system are given by

 
\begin{subequations}
	\begin{align}
		\label{Eq:secondkuramoto1n} M\frac{d^{2}\theta_{1}(t)}{{dt}^2}+\alpha\frac{d\theta_{1}(t)}{dt} =I_{1}-K^{'} 
		\sin\left(\theta_{1}(t)-\theta_{2}(t-\tau)\right),\\ 
		\label{Eq:secondkuramoto2n}M\frac{d^{2}\theta_{2}(t)}{{dt}^2}+\alpha\frac{d\theta_{2}(t)}{dt} =I_{2}-K^{'} 
		\sin\left(\theta_{2}(t)-\theta_{1}(t-\tau)\right),
	\end{align}
\end{subequations} 
where $ \theta_{i} $  represents the phase of $i$th oscillator. $K^{'}$, $\tau$, $\alpha$, denote the coupling strength, the time delay, and the damping coefficient, respectively. Dividing Eqs.~\eqref{Eq:secondkuramoto1n} and ~\eqref{Eq:secondkuramoto2n} by the damping coefficient,  one obtains the following equations:

\begin{subequations}
	\begin{align}
		\label{Eq:secondkuramoto1} m\frac{d^{2}\theta_{1}(t)}{{dt}^2}+\frac{d\theta_{1}(t)}{dt} =\omega_{1}-K 
		\sin\left(\theta_{1}(t)-\theta_{2}(t-\tau)\right),\\ 
		\label{Eq:secondkuramoto2}m\frac{d^{2}\theta_{2}(t)}{{dt}^2}+\frac{d\theta_{2}(t)}{dt} =\omega_{2}-K 
		\sin\left(\theta_{2}(t)-\theta_{1}(t-\tau)\right),
	\end{align}
\end{subequations} 
where $m=\frac{M}{\alpha}$, $K=\frac{K^{'}}{\alpha}$, $\omega_{1}=\frac{I_{1}}{\alpha}$, and $\omega_{2}=\frac{I_{2}}{\alpha}$.

The Eqs.~\ref{Eq:secondkuramoto1} and ~\ref{Eq:secondkuramoto2} represent the standard form of the second-order Kuramoto model~\cite{tanaka1997self}. In this form, $\omega_{1}$ and $\omega_{2}$ correspond to the intrinsic frequencies of the first-order Kuramoto model.
The phase-locked state occurs when two oscillators have a constant phase difference ($\beta$) over time. Therefore, the phase of the oscillators in the stationary state is given as follows:
\begin{eqnarray}
	\theta_{1,2}(t)=\varphi(t)\pm \frac{\beta}{2}.
	\label{Eq:definephase} 
	\end{eqnarray}
	By substituting Eq.~\eqref{Eq:definephase} into Eqs.~\eqref{Eq:secondkuramoto1} and ~\eqref{Eq:secondkuramoto2} and subtracting the resulting equations from each other, the following condition is derived:

\begin{eqnarray} 
	\cos(\varphi(t)-\varphi(t-\tau))=\frac{\omega_{1}-\omega_{2}}{2K\sin{\beta}}={\rm const},
	\label{Eq:condition1} \end{eqnarray} 
	which makes us choose $\varphi(t)=\Omega t$, where $\Omega$ refers to the final frequency of the oscillators in the phase-locked state. By employing the newly defined oscillator phases and utilizing the equation of the second-order Kuramoto model, we obtain two new equations:

\begin{subequations}
	\begin{align}
		\label{Eq:eq1} \Delta\omega =2K\sin{\beta} \cos(\Omega\tau),\\ 
		\label{Eq:eq2} \Omega=\bar{\omega}-K\sin(\Omega\tau)\cos{\beta}.
	\end{align}
\end{subequations}
here, the terms $\Delta\omega=(\omega_1 - \omega_2)$ and $\bar{\omega}=\frac{\omega_1 + \omega_2}{2}$ denote the difference and average of intrinsic frequencies, respectively.
Phase-locked solutions are obtained by determining the values of $\Omega$ and $\beta$. Elimination of  phase difference $\beta$ from the above equations, one can determine  the final frequency $\Omega$ by finding  zeros of the following equations:

\begin{subequations}
	\begin{align}
		\label{Eq:func1}  f_{-}(\Omega) =\bar{\omega}-\Omega - K \tan(\Omega\tau) \sqrt{\cos^2 (\Omega\tau)-{\biggl(\frac{\Delta\omega}{2k}}\biggr)^2},\\ 
		\label{Eq:func2} f_{+}(\Omega) =\bar{\omega}-\Omega + K \tan(\Omega\tau) \sqrt{\cos^2 (\Omega\tau)-{\biggl(\frac{\Delta\omega}{2k}}\biggr)^2}.
	\end{align}
\end{subequations}The phase difference between oscillators is defined using Eq.~\eqref{Eq:eq1} as follows:

\begin{eqnarray} 
	\beta_{f_{-}(\Omega)} =
	\begin{cases}
		\arcsin(\frac{\Delta\omega}{2K\cos(\Omega\tau)}) & \text{if $\cos(\Omega\tau)>0$} \\\\
		\pi+\arcsin(\frac{\Delta\omega}{2K|\cos(\Omega\tau)|}) & \text{otherwise}
	\end{cases}
	\label{Eq:beta1} \end{eqnarray}

\begin{eqnarray} 
	\beta_{f_{+}(\Omega)} =
	\begin{cases}
		\pi-\arcsin(\frac{\Delta\omega}{2K\cos(\Omega\tau)}) & \text{if $\cos(\Omega\tau)>0$} \\\\
		-\arcsin(\frac{\Delta\omega}{2K|\cos(\Omega\tau)|}) & \text{otherwise.}
	\end{cases}
	\label{Eq:beta2} 
	\end{eqnarray}
	
We use the linear stability analysis to examine the solutions' stability. This involves introducing a small disturbance to each phase and analyzing the resulting dynamics~\cite{driver2012ordinary}.
The perturbed phases can be expressed as:

\begin{eqnarray}
\theta_{1}(t)=\Omega t+\frac{\beta}{2}+a e^{\lambda t} \nonumber\\
 \theta_{2}(t)=\Omega t-\frac{\beta}{2}+b e^{\lambda t},\nonumber
\end{eqnarray} 
 and by inserting the above ansatz in Eqs.~\eqref{Eq:secondkuramoto1} and ~\eqref{Eq:secondkuramoto2} and linearizing the equations, we can determine the exponent $\lambda$   by finding the zeros of  the following determinant:

\begin{eqnarray} 
	\begin{vmatrix}
		m{\lambda}^2+\lambda+K \cos(\Omega\tau+\beta) & -K e^{-\lambda\tau} \cos(\Omega\tau+\beta)\\
		-K e^{-\lambda\tau} \cos(\Omega\tau-\beta)&m{\lambda}^2+\lambda+K \cos(\Omega\tau-\beta).
	\end{vmatrix} 
	\label{Eq:lambda}
	 \end{eqnarray}
The exponent $\lambda$ indicates the stability of the solutions. For $Re(\lambda)<0$, the resulting solution is stable, otherwise, it is unstable. 
	
	 Eqs.~\eqref{Eq:func1} to ~\eqref{Eq:beta2} are independent of $m$, and are the same obtained for the first order Kuramoto model~\cite{ schuster1989mutual}. However, Eq.~\eqref{Eq:lambda} implies that the presence of inertia in the second-order model could affect the stability of solutions.

To measure the degree of  synchrony between the two oscillators we use the Kuramoto  order parameter ($r$) defined  as: 
\begin{eqnarray}
	r (t)& = & \frac{1}{2} \mid \sum_{j=1}^2 e^{i\theta_{j}(t)}\mid=\mid\cos(\frac{\theta_1-\theta_2}{2})\mid,
	\label{Eq:r}
\end{eqnarray} 
 {If the two oscillators are in-phase ($\theta_1-\theta_2=0$)  the order parameter is equal to unity, and it is equal to zero if they are in opposite phase ($\theta_1-\theta_2=\pi$) }.

\section{Results and discussion}
\label{results}

\subsection{Case $\omega_1=\omega_2$}
\begin{figure*}[]
	\centering
	 \subfigure[\label{a}]{\includegraphics[width=0.5\columnwidth]{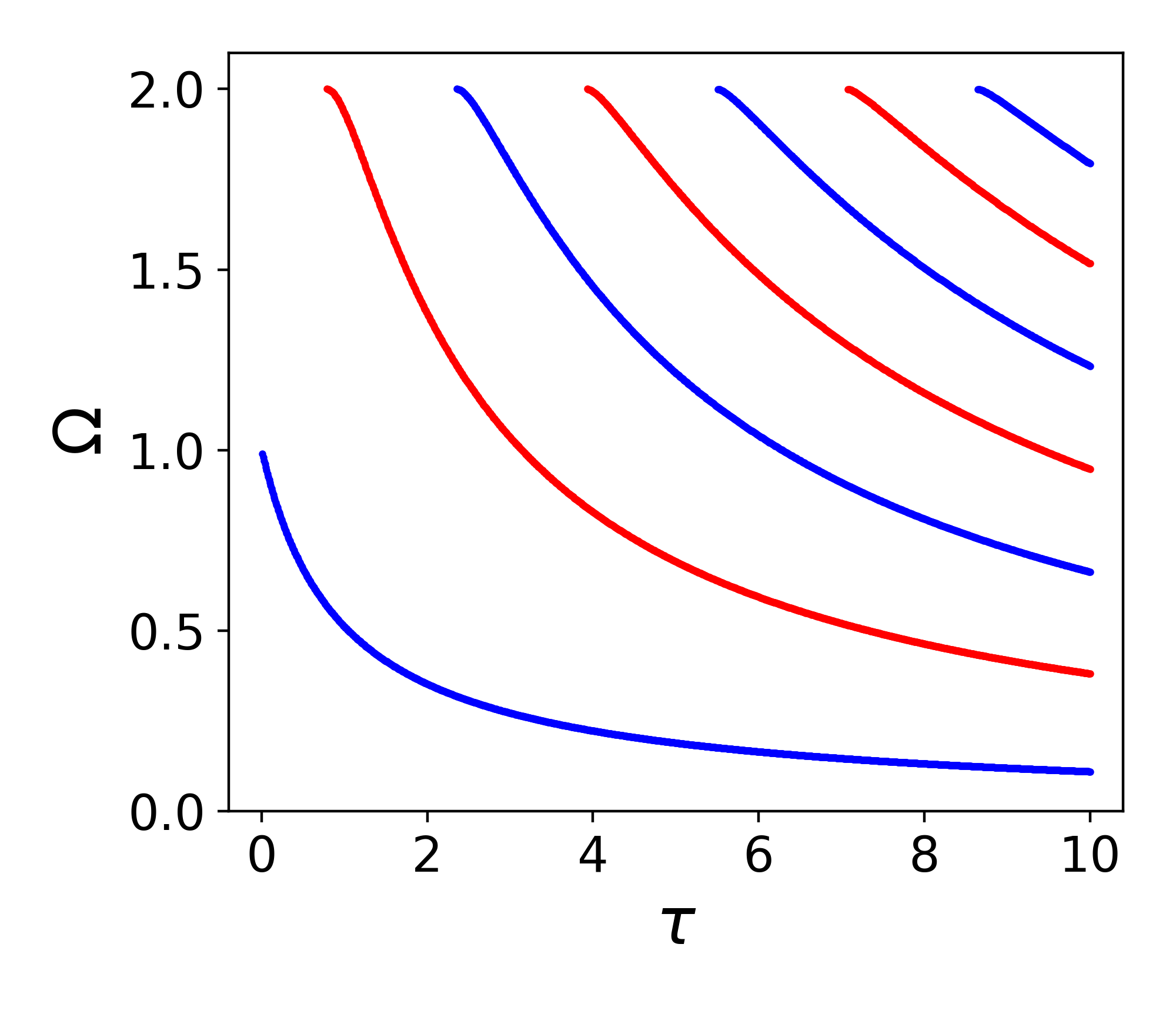}}
	 \subfigure[\label{b}]{\includegraphics[width=0.5\columnwidth]{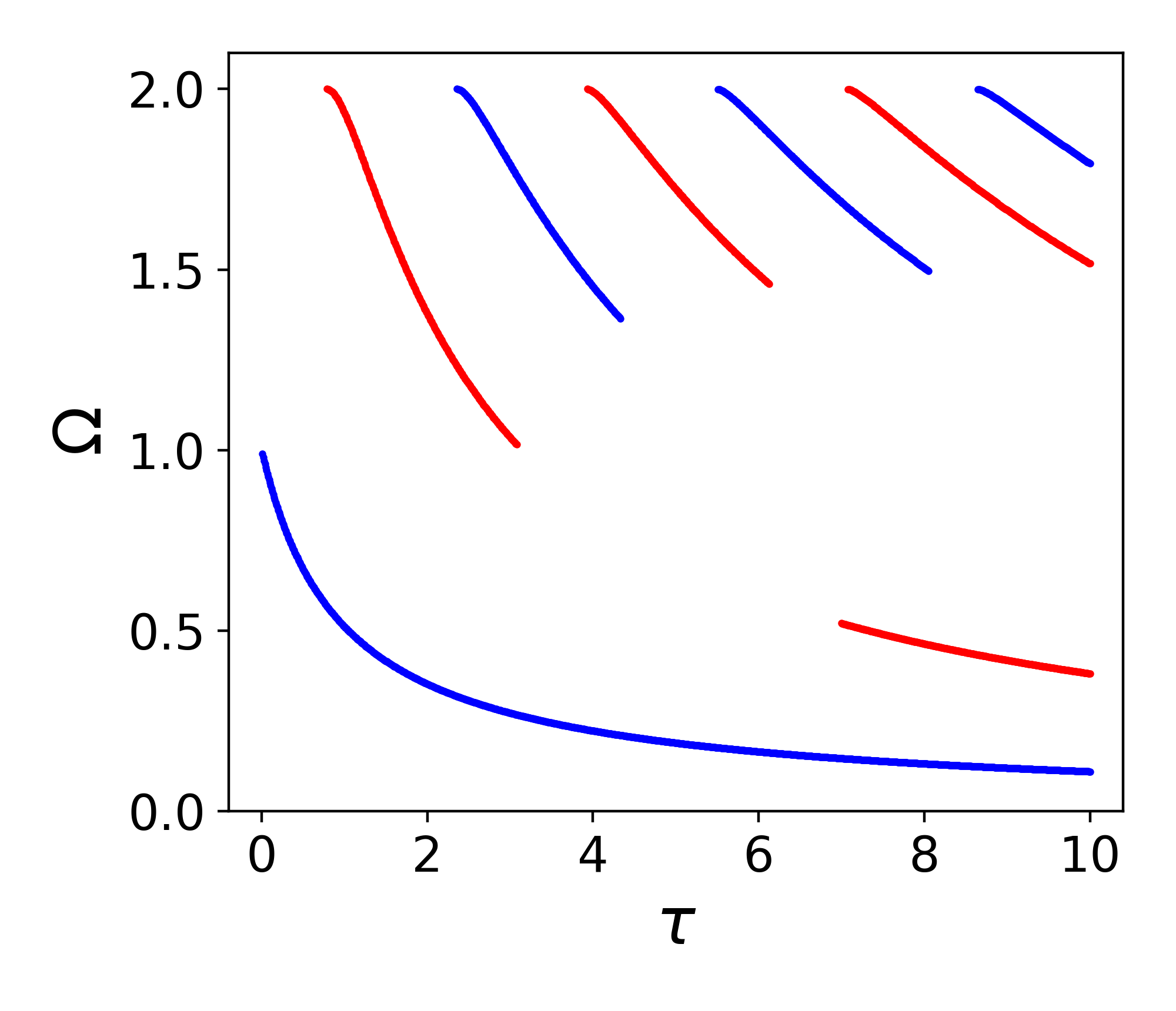}}
	 \subfigure[\label{c}]{\includegraphics[width=0.5\columnwidth]{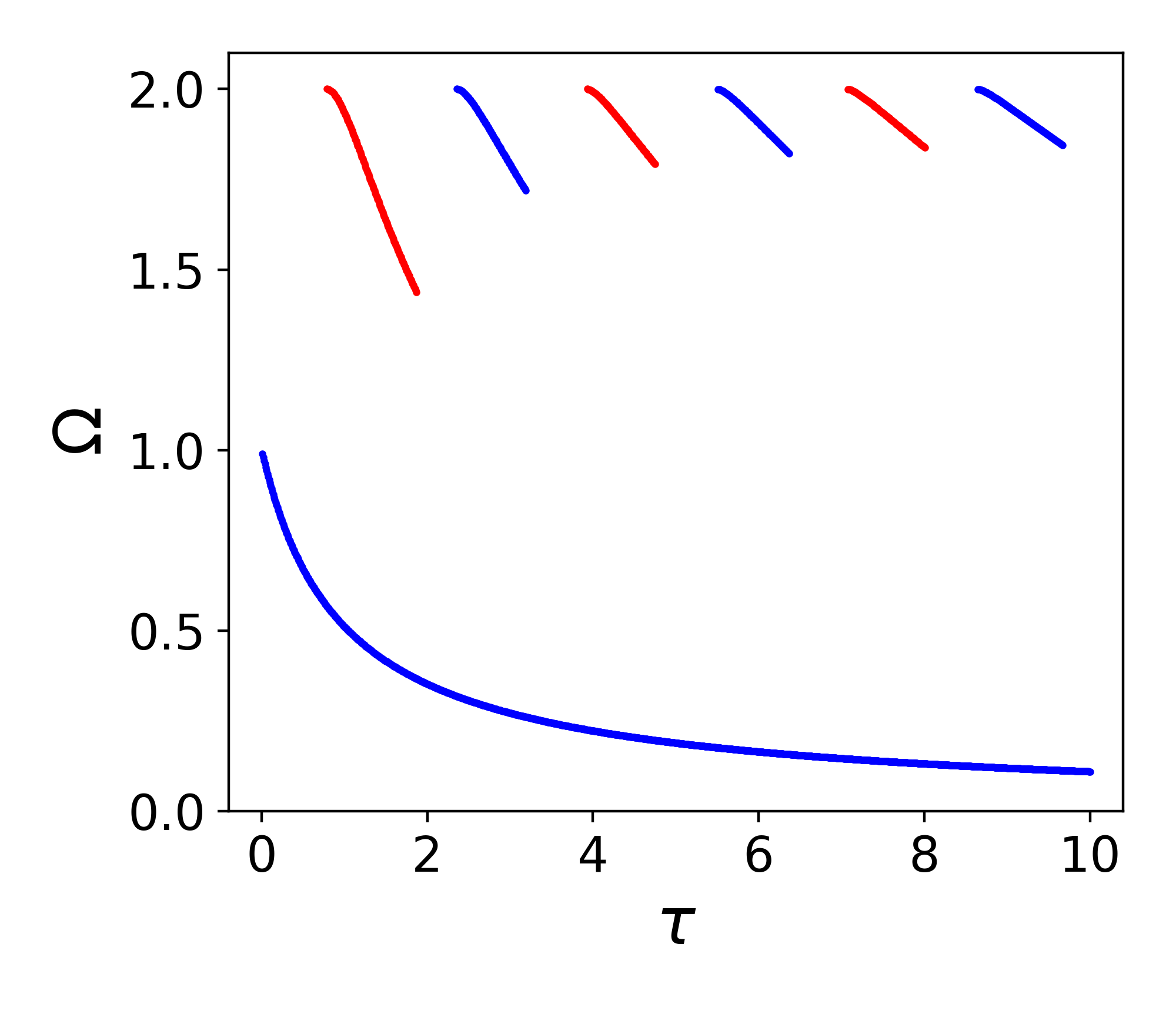}}
	 \subfigure[\label{d}]{\includegraphics[width=0.5\columnwidth]{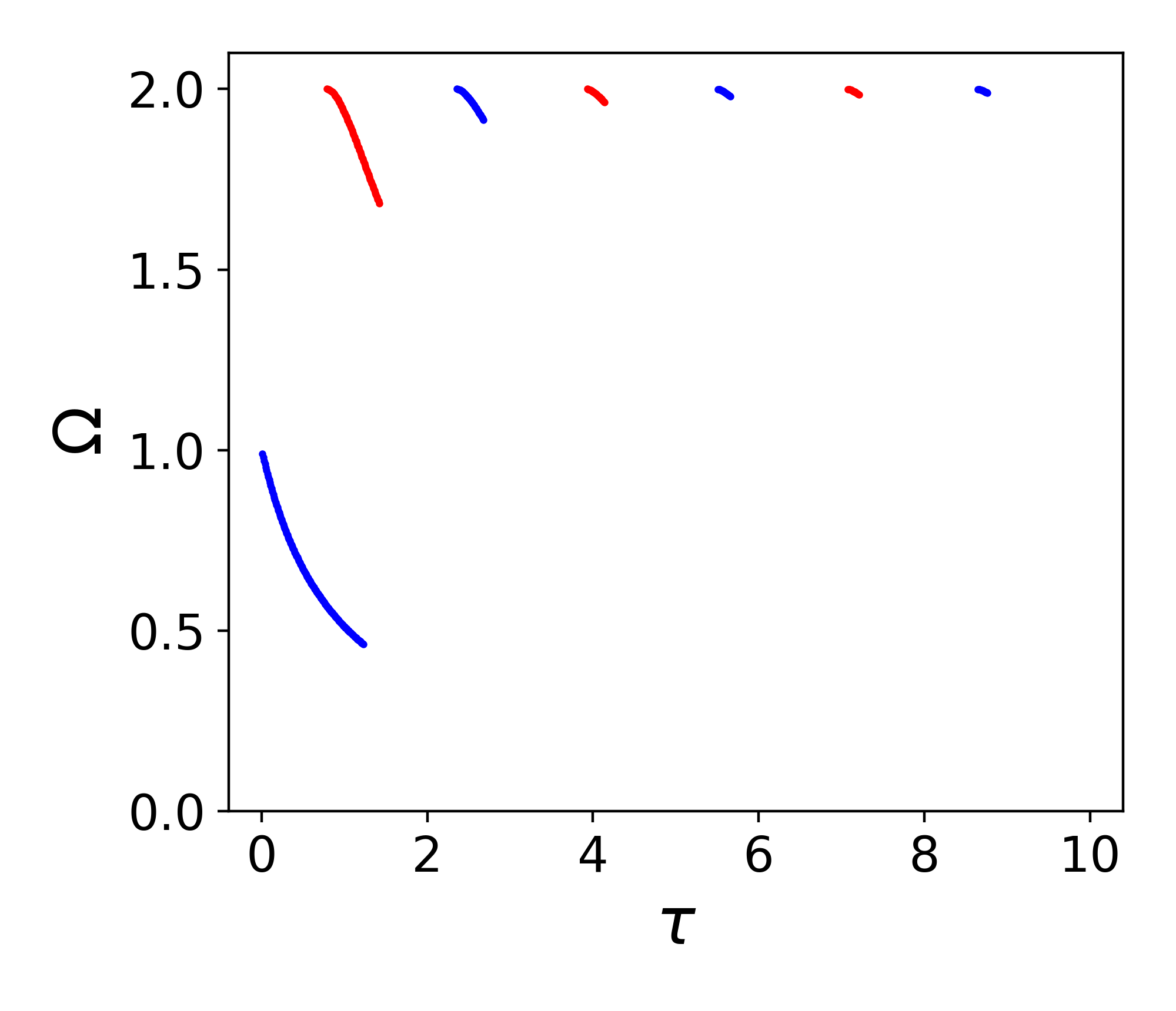}}	
	\caption{Stable phase-locked  frequencies of the oscillators versus time delay for  $K=1$, $\omega_1=\omega_2=1$ and  various values of $m$, including (a) $m=0$, (b) $m=0.6$, (c) $m=1.0$ and (d) $m=10$. Blue and red dots represent the stable phase-locked solutions with the phase difference $\beta=0$ and $\beta=\pi$, respectively.}
	\label{fig1:w-tau}
\end{figure*}

\begin{figure*}[]
	\centering
	\subfigure[\label{a}]{\includegraphics[width=0.5\columnwidth]{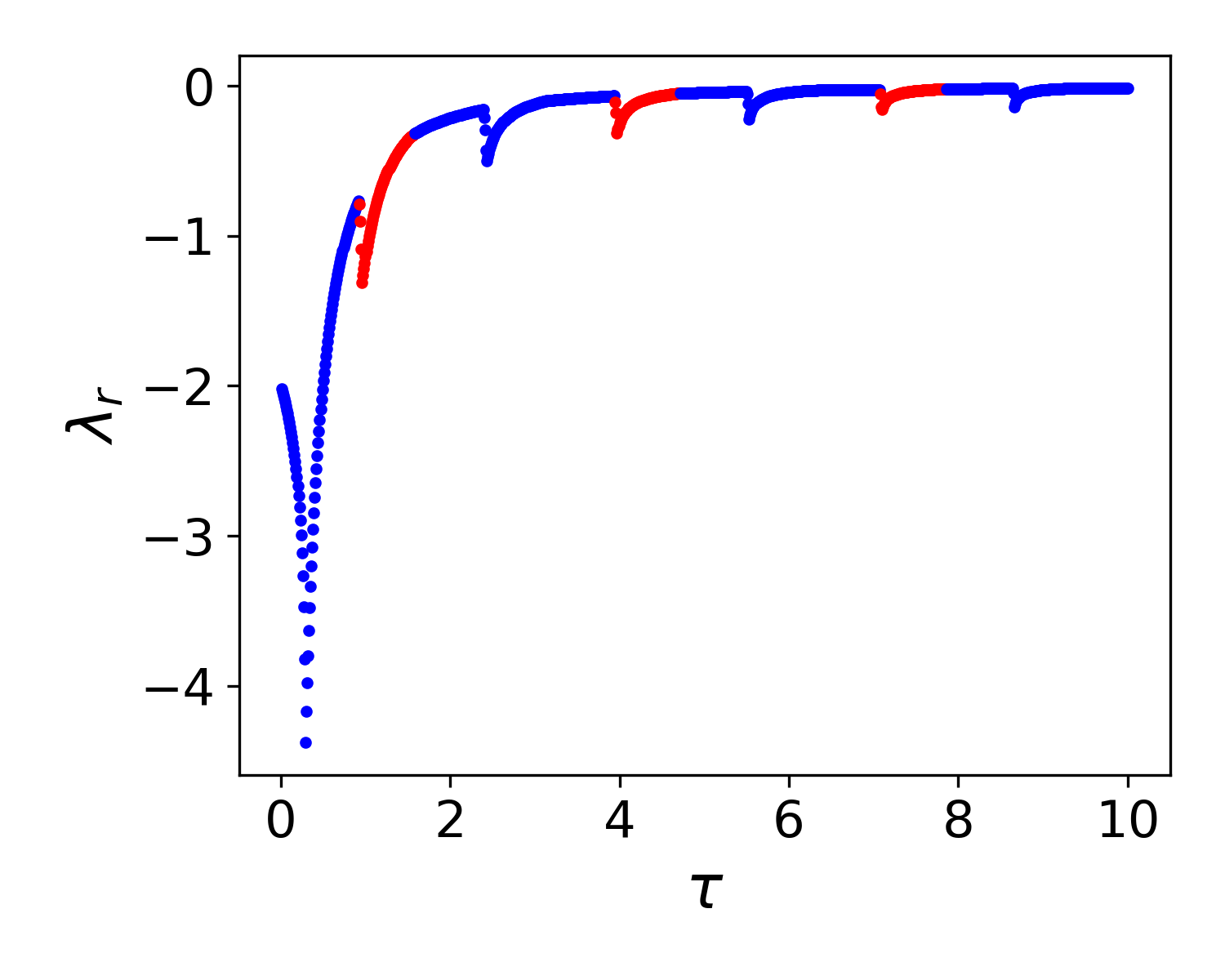}}
	 \subfigure[\label{b}]{\includegraphics[width=0.5\columnwidth]{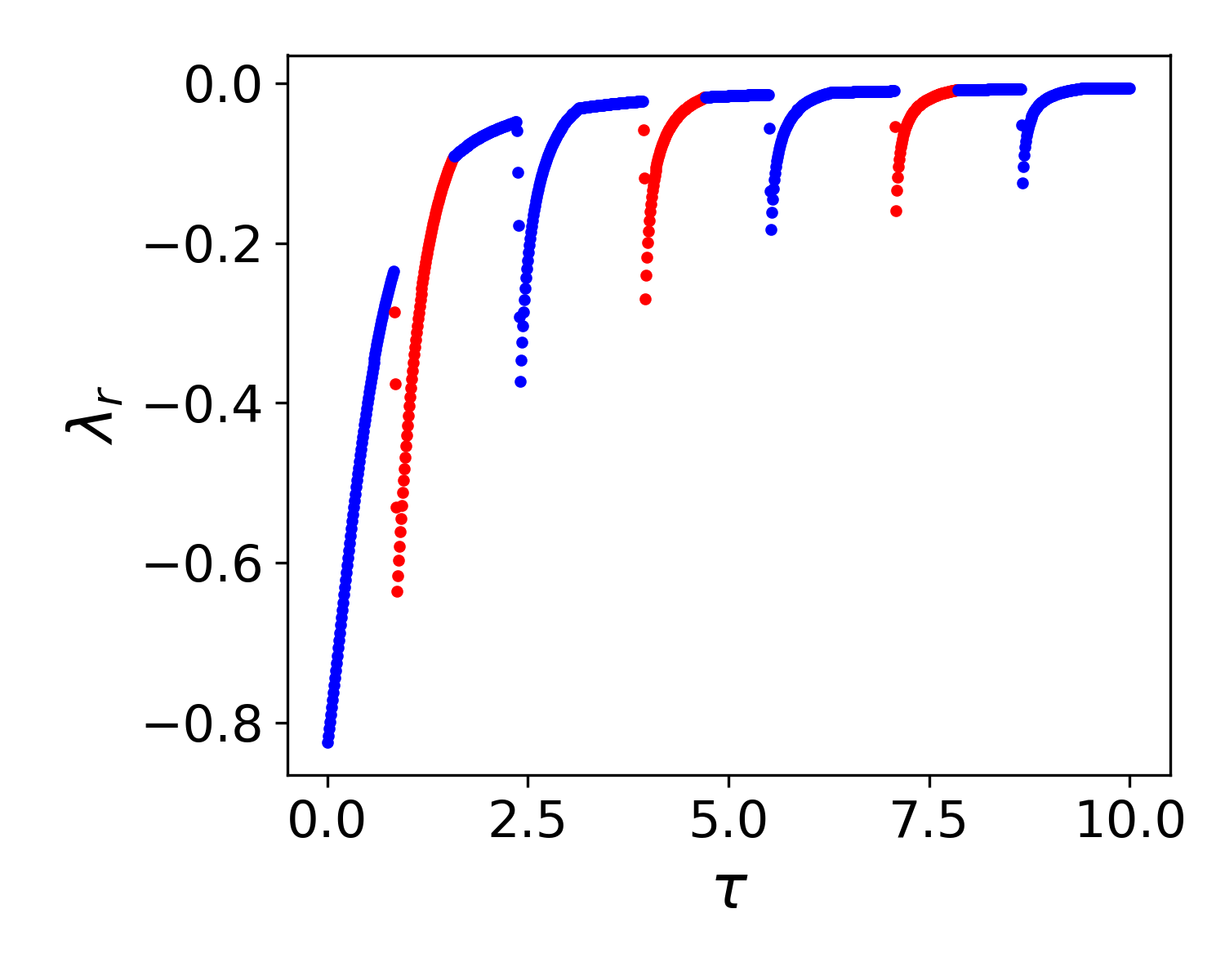}}
	 \subfigure[\label{c}]{\includegraphics[width=0.5\columnwidth]{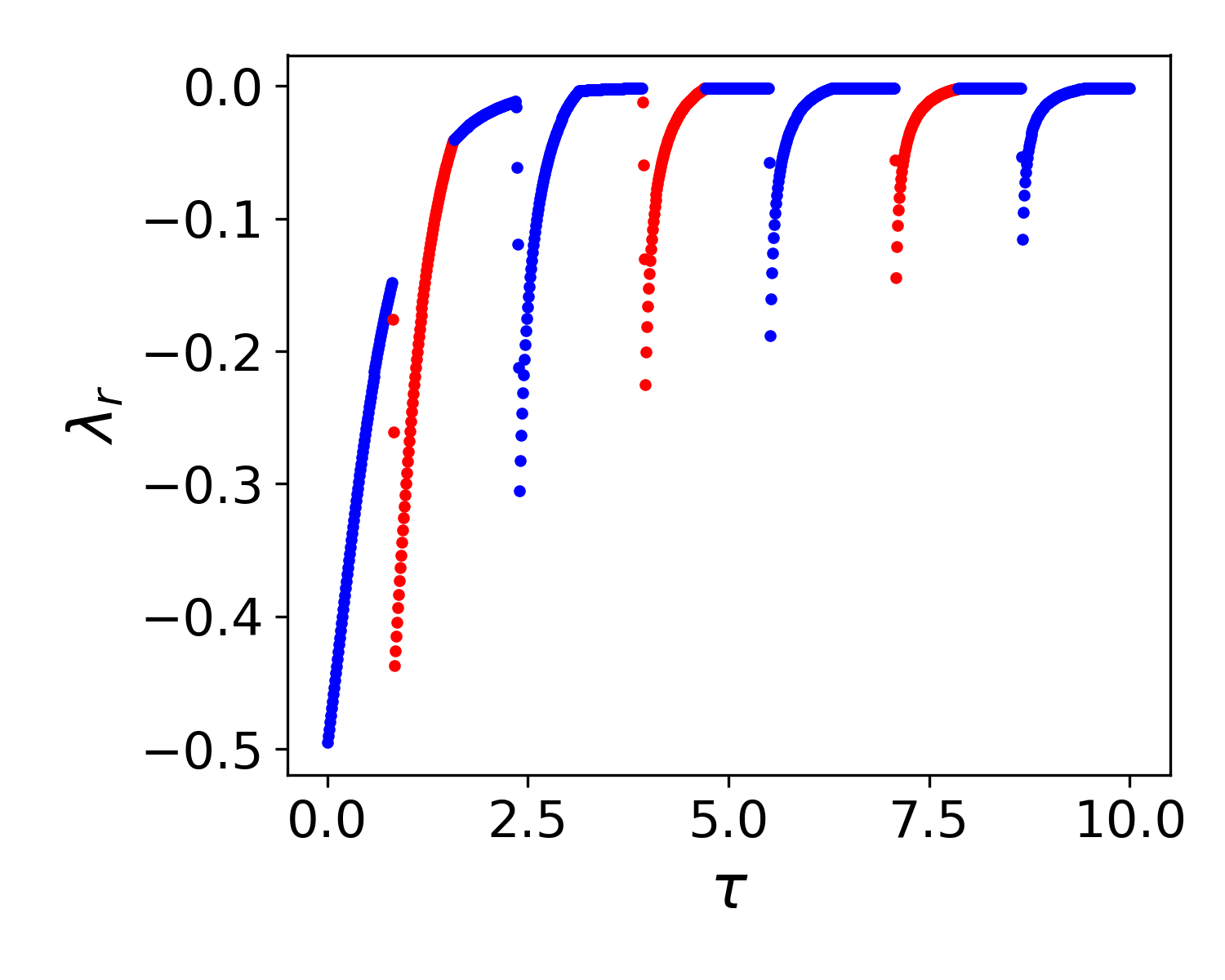}}
	 \subfigure[\label{d}]{\includegraphics[width=0.5\columnwidth]{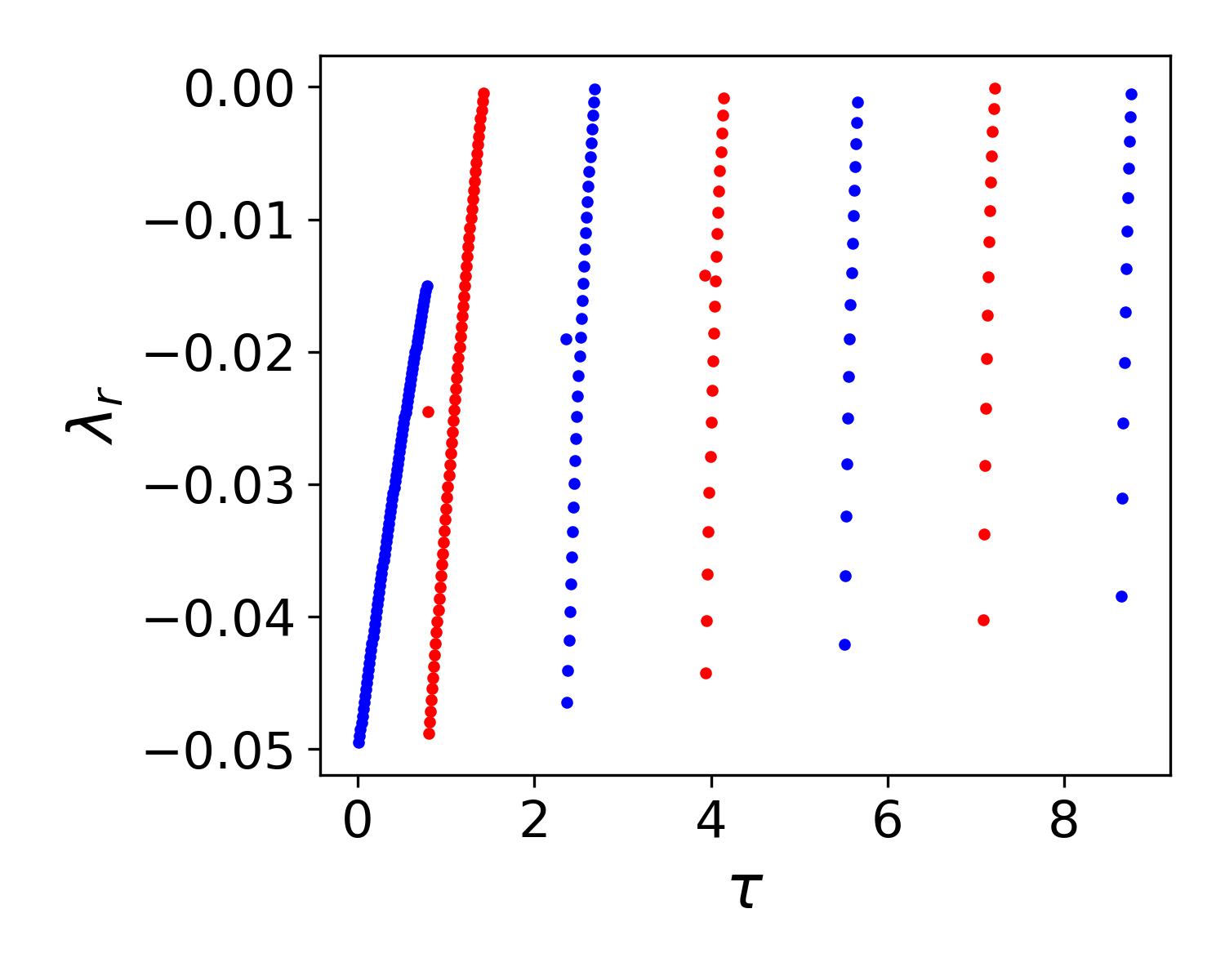}}	
	\caption{The real part of $\lambda$ exponent corresponding to the most stable phase-locked frequency versus time delay, considering $K=1$, $\omega_1=\omega_2=1$, and  (a) $m=0$, (b) $m=0.6$, (c) $m=1.0$, and (d) $m=10$. Blue and red dots represent the stable phase-locked solutions with the phase difference $\beta=0$ and $\beta=\pi$, respectively.}
	\label{fig2:Lyapunov}
\end{figure*}

\begin{figure*}[]
	\centering
	\subfigure[\label{a}]{\includegraphics[width=0.5\columnwidth]{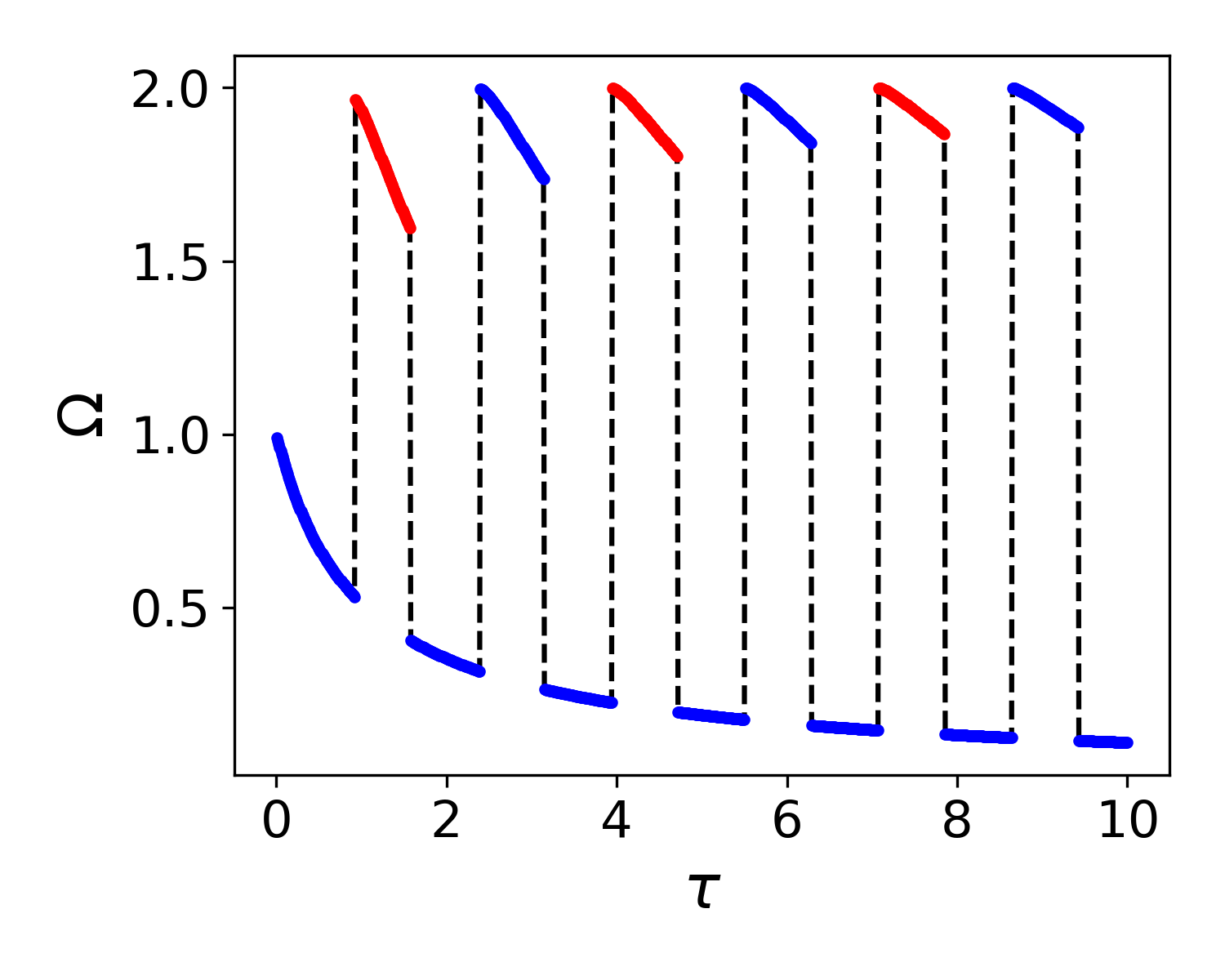}}
	 \subfigure[\label{b}]{\includegraphics[width=0.5\columnwidth]{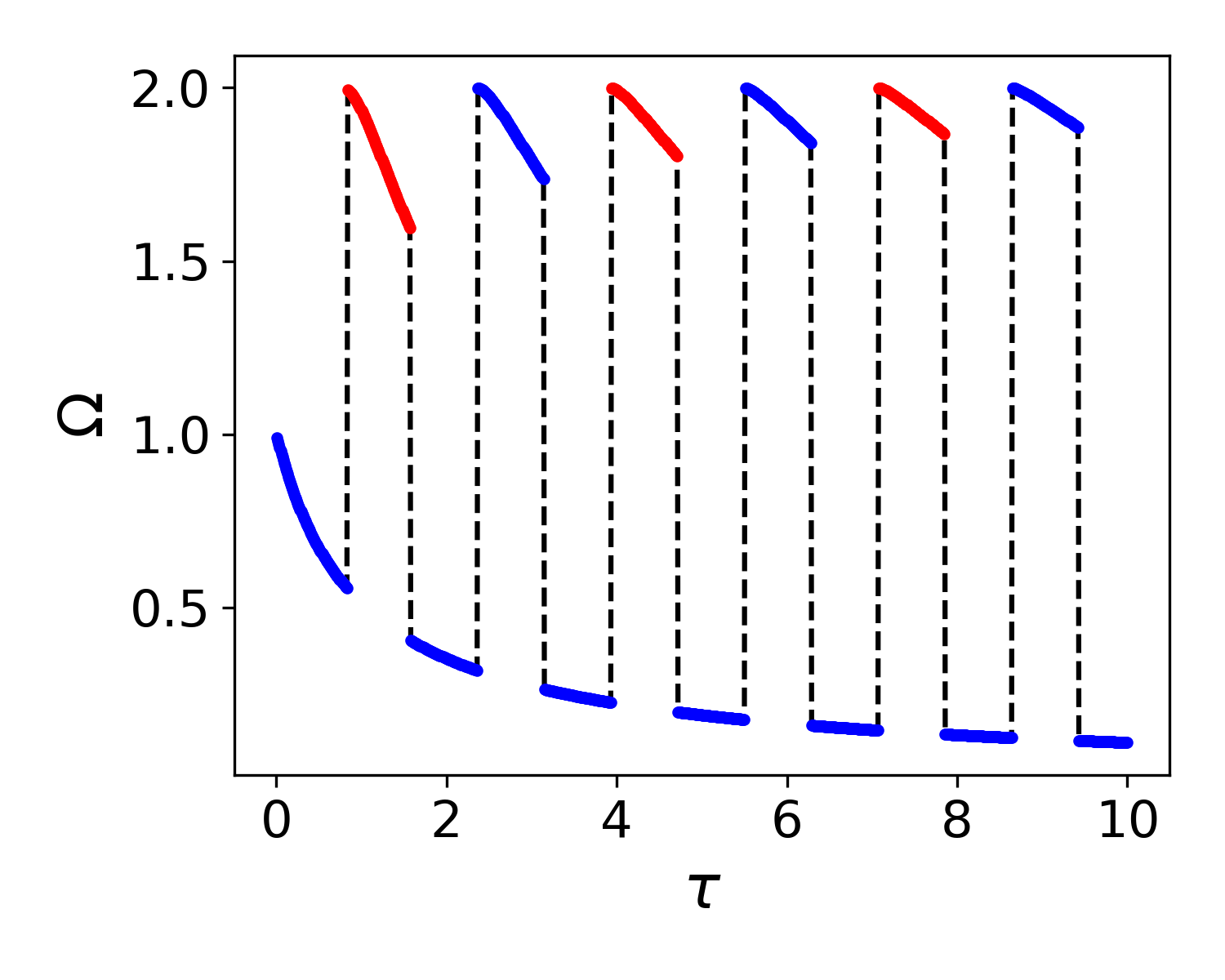}}
	 \subfigure[\label{c}]{\includegraphics[width=0.5\columnwidth]{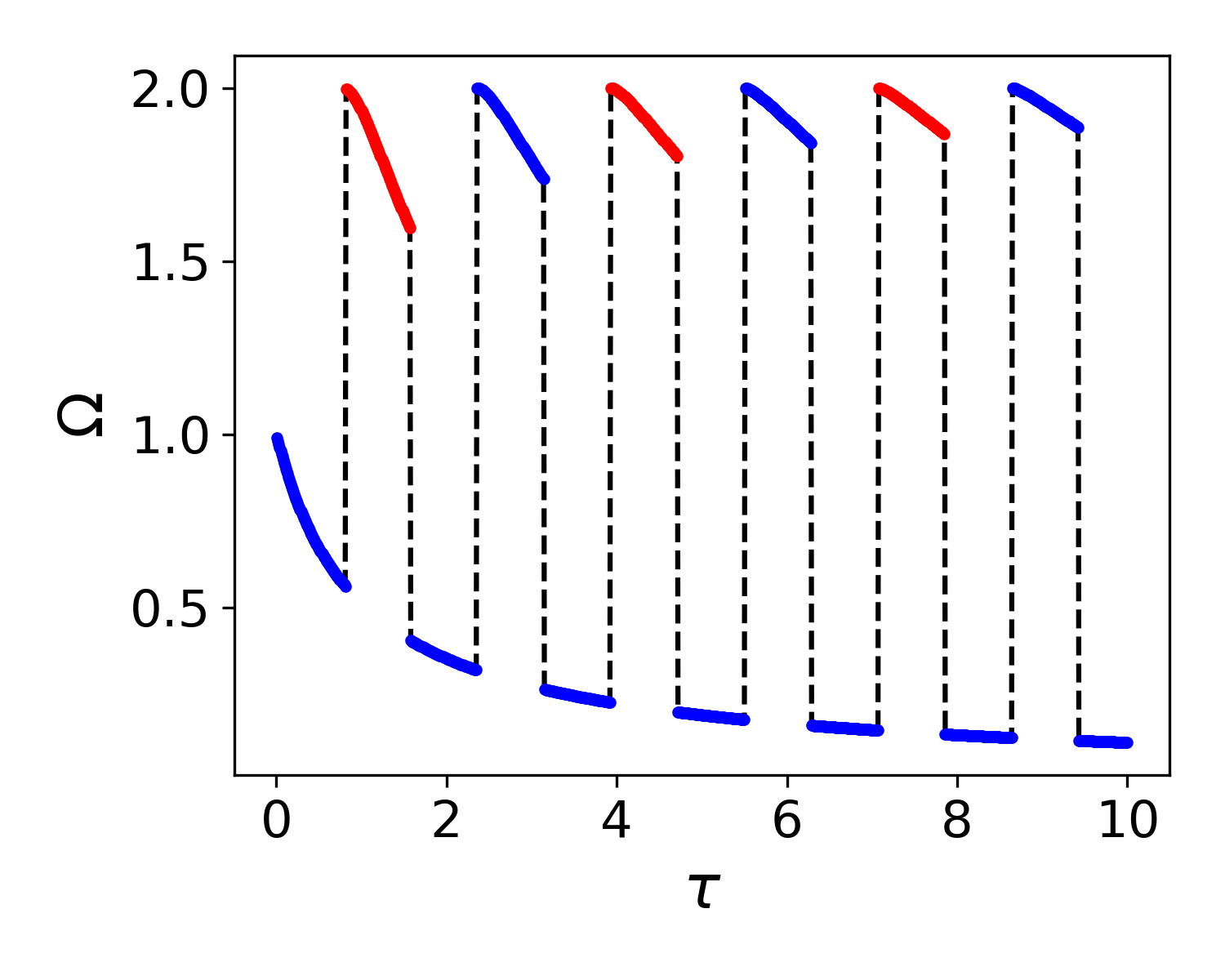}}
	 \subfigure[\label{d}]{\includegraphics[width=0.5\columnwidth]{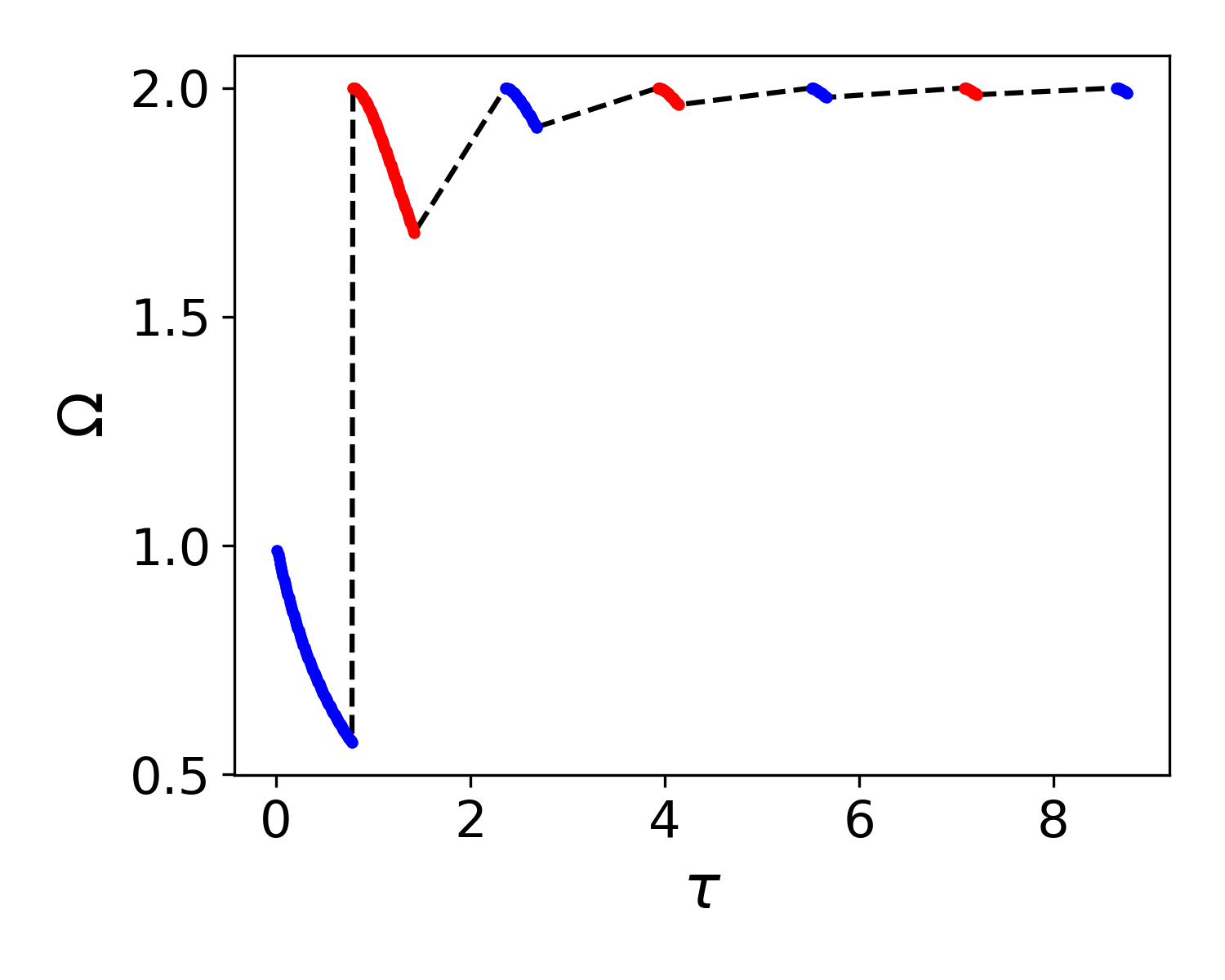}}	
	\caption{The most stable phase-locked frequencies versus time delay, considering $K=1$, $\omega_1=\omega_2=1$, and  (a) $m=0$, 
	(b) $m=0.6$, (c) $m=1.0$, (d) $m=10$. Blue and red dots represent the stable phase-locked solutions with the phase difference $\beta=0$ 
	and $\beta=\pi$, respectively.}
	\label{fig3:most-st}
\end{figure*}

\begin{figure*}[]
	\centering
	\subfigure[\label{a}]{\includegraphics[width=0.5\columnwidth]{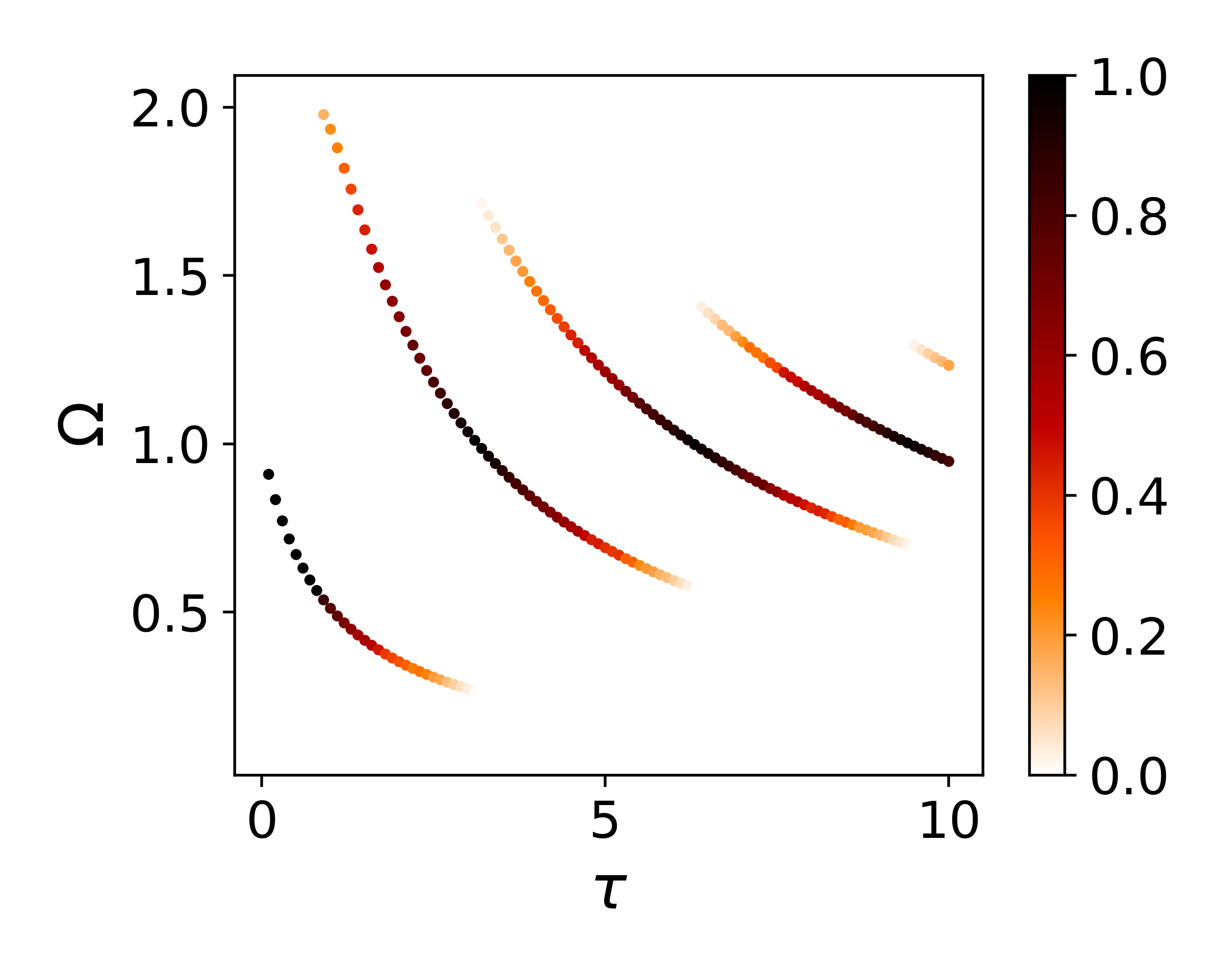}}
	 \subfigure[\label{b}]{\includegraphics[width=0.5\columnwidth]{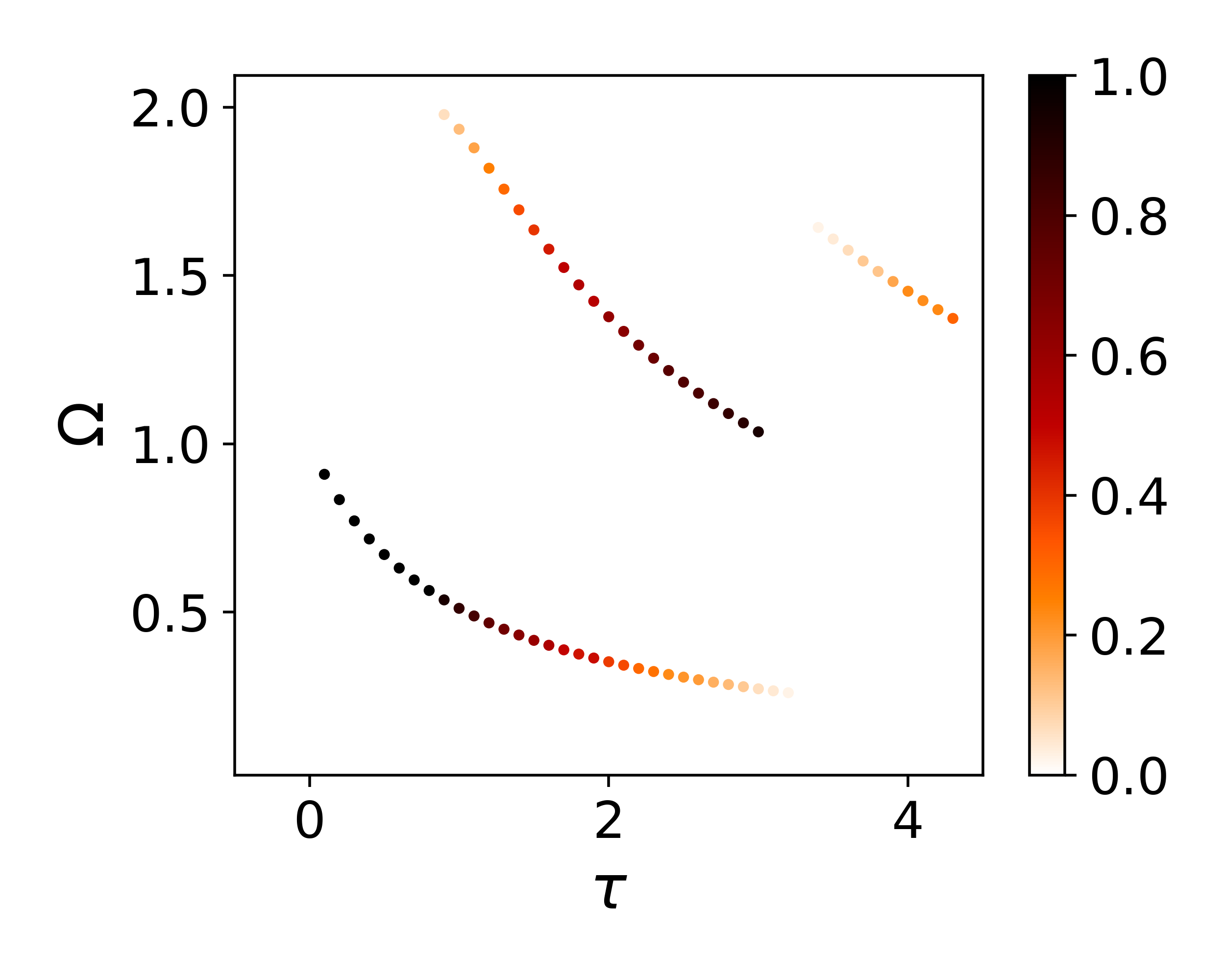}}
	 \subfigure[\label{c}]{\includegraphics[width=0.5\columnwidth]{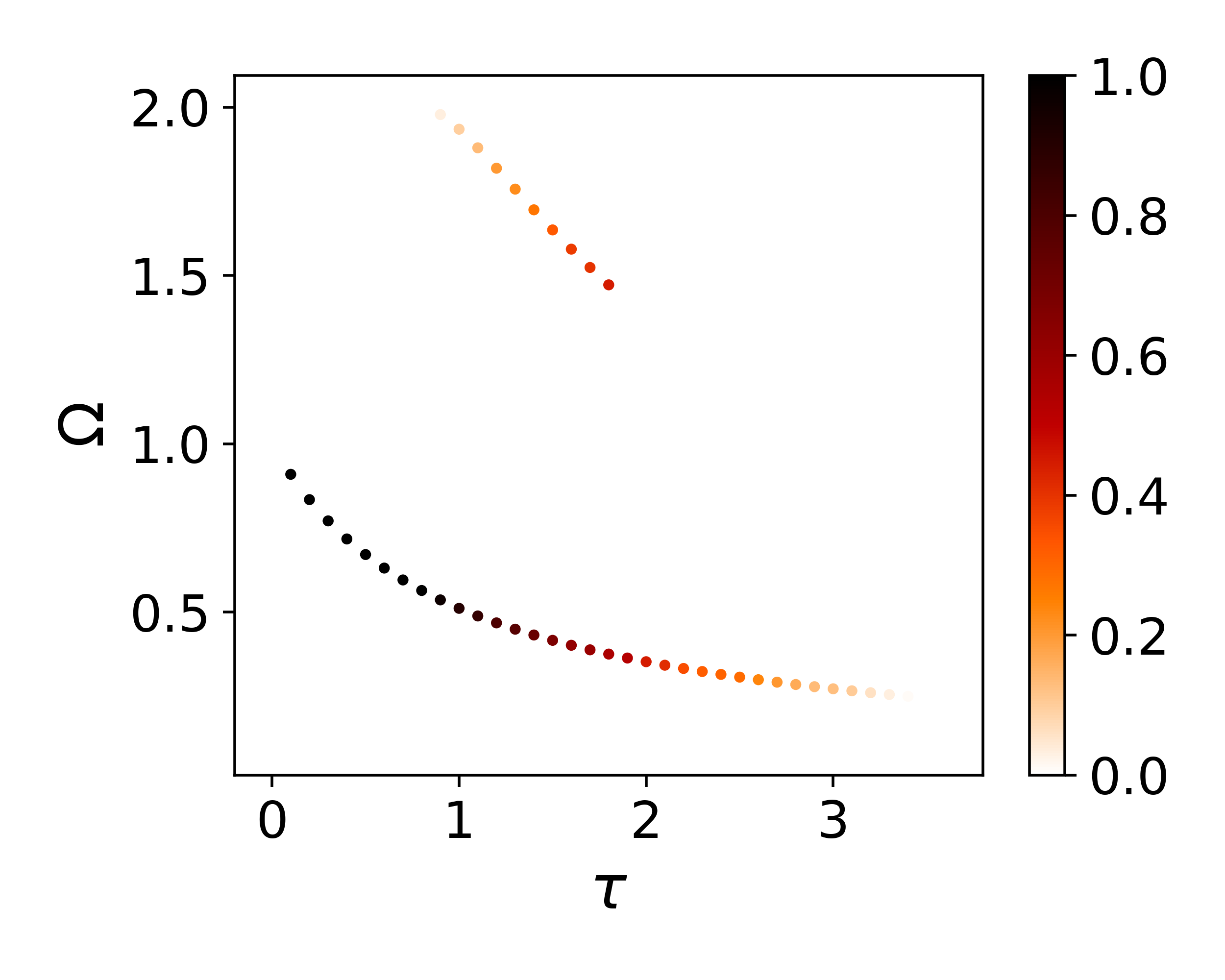}}
	 \subfigure[\label{d}]{\includegraphics[width=0.5\columnwidth]{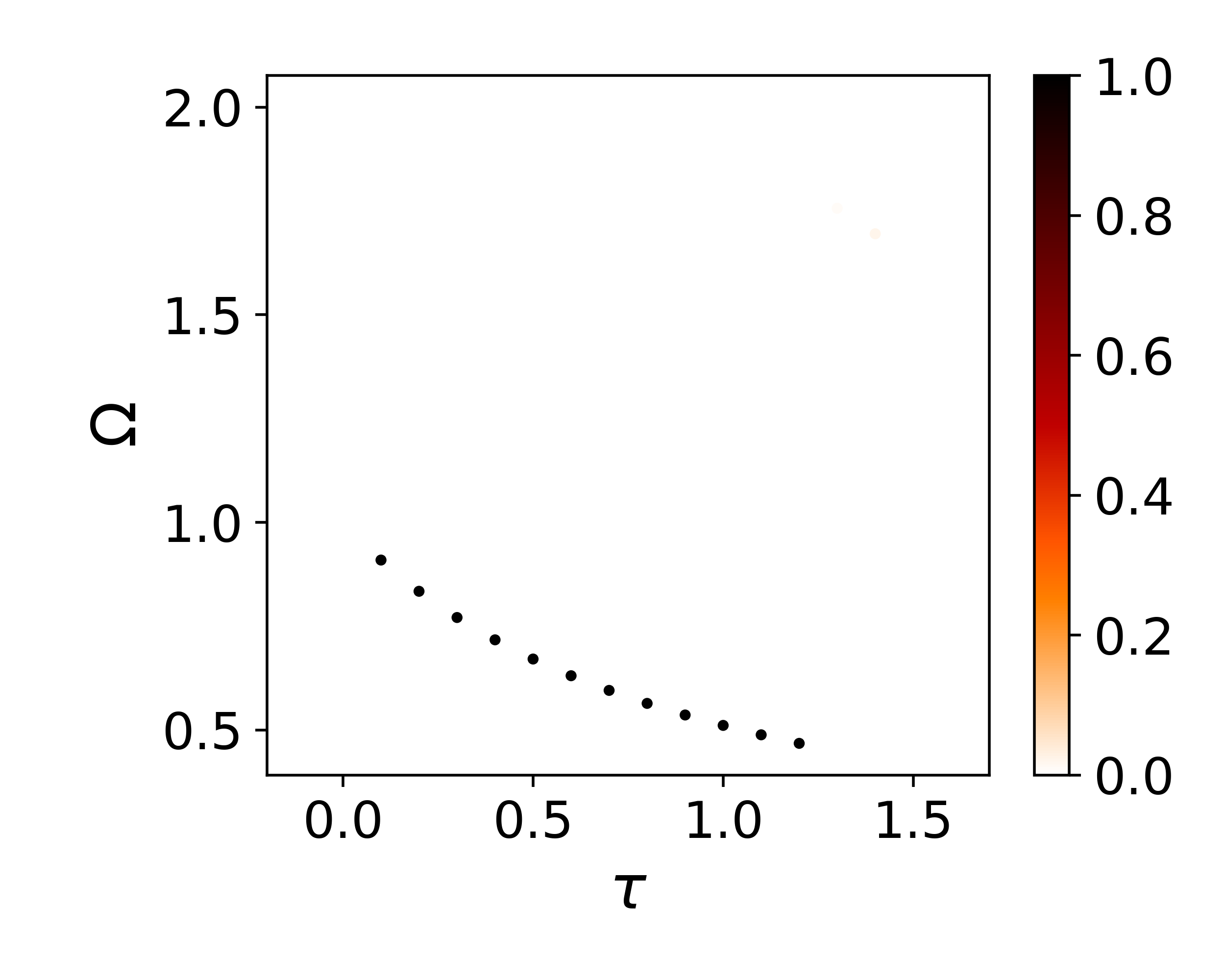}}	
	\caption{Probability of phase-locked solutions estimated over 500 realizations for $K=1$ and $\omega_1=\omega_2=1$, and  
	(a) $m=0$, (b) $m=0.6$, (c) $m=1.0$, (d) $m=10$. The probability of each solution is encoded by a color with a value indicated in the 
	     sidebar.}
	\label{fig4:numeric1}
\end{figure*}

To simplify our analysis of phase-locked solutions, we will consider a specific case where both oscillators are identical. This means that their masses and intrinsic frequencies are the same, hence, we will set them at $\omega_1=\omega_2=1$. Moreover, we will assume that $K=1$. When the coupling strength is positive ($K^{'}>0$), all solutions obtained through $f_{+}(\Omega)$ are unstable, regardless of the value of $m$. Therefore, we will only focus on exploring solutions of $f_{-}(\Omega)$ in our subsequent analysis.

We started by investigating the stability of the solutions of Equation \eqref{Eq:func1} by finding the zeros of the determinant \eqref{Eq:lambda}. 
We found that this equation typically does not have only real solutions. Therefore, we assumed that the $\lambda$ exponents are complex numbers. The stability of the fixed point solutions requires that the real part of the largest $\lambda$ exponent becomes negative.
Fig.~\ref{fig1:w-tau}, illustrates the stable phase-locked solutions $\Omega$ as a function of the time delay ($\tau$) for $m=0$, $0.6$, $1$, and $10$. Meanwhile, Figs.~\ref{fig2:Lyapunov} and \ref{fig3:most-st} show the delay dependence of the real part of $\lambda$ exponents and locking frequencies corresponding to the most stable solutions, respectively. The blue dots represent solutions associated with the phase difference $\beta=0$, while the red dots represent solutions associated with $\beta=\pi$. 

Fig.~\ref{fig1:w-tau} displays that for values of $\tau$ up to 0.79, there is only one stable solution in which the system is phase-locked, and the frequency of the system decreases as the time delay increases. However, at $\tau = 0.79$, the system becomes bistable, meaning that there are now two stable phase-locked solutions. The figure further indicates that the larger values of time delay result in multistability. It is important to note that in the absence of inertia,  phase-locked solutions exist for any value of time delay. However, considering the second-order Kuramoto model ($m \neq 0$), the presence of inertia can cause the phase-locked solutions to become unstable. As the value of $m$ increases, the range of instabilities expands, and the system becomes less stable. For instance, when $m = 10$, only a small part of the phase-locked branches remains stable.

To determine the basin of attraction for each phase-locked solution, we conducted numerical integration on equations \ref{Eq:secondkuramoto1} and \ref{Eq:secondkuramoto2}. For this purpose, we employed the fourth-order Runge-Kutta method with a time step of $dt=0.01$. We also checked our results using the 6th-order predictor-corrector technique and discovered no significant variances. The initial phases of the two oscillators were chosen randomly from a uniform distribution within the range of $[-\pi, \pi[$, with the initial angular velocities set to zero. The oscillators were allowed to evolve independently in the time interval of $[0, \tau]$ before being made to interact with each other for $t > \tau$. The size of the basin of attraction associated with each fixed point was estimated as the probability of converging any initial condition to that fixed point over 500 realizations of simulation. 

The results of numerical integration for $m=0, 0.6, 1$, and $10$ are displayed in Fig.~\ref{fig4:numeric1}. The figure shows the probability of each phase-locked solution through a color code that is indicated on the side. The color code reveals that the most likely solutions are the ones with a frequency close to the average value $\bar{\omega}=1$. For each time delay, solutions with a locking frequency farther from 1 have a smaller basin of attraction. Interestingly, the most stable solutions do not necessarily have the largest basin of attraction. Higher inertia makes finding solutions on higher branches and larger time delays less probable. For instance, Fig.~\ref{fig4:numeric1}-(d) shows that for $m=10$, only the lowest branch is observed in numerics for $\tau\lesssim 1.2$.

\begin{figure*}[]
	\centering
	\renewcommand\thesubfigure{\fontsize{11}{11}\selectfont (\alph{subfigure})}
	\fbox{\subfigure[\label{aa}]{\includegraphics[width=0.45\textwidth]{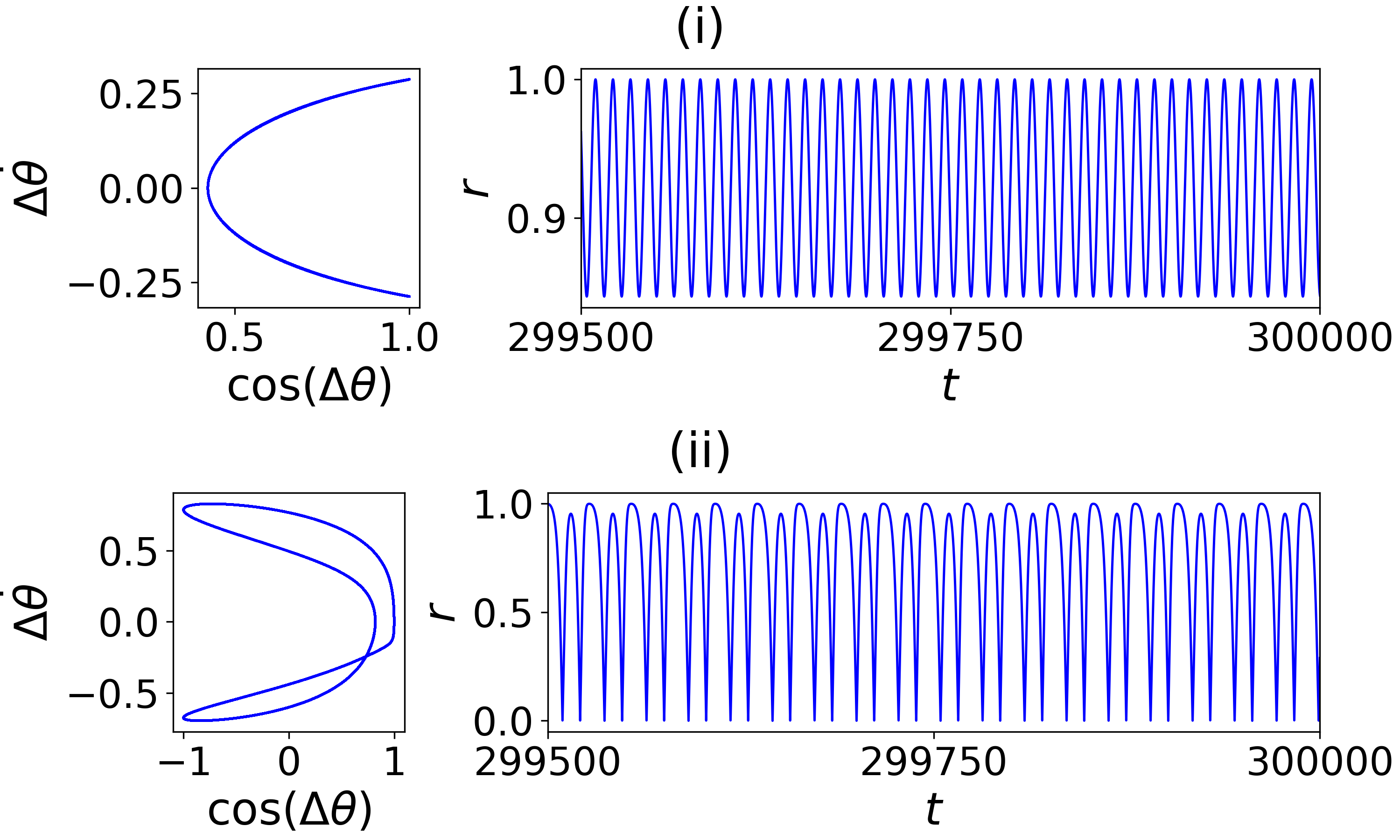}}}
	\fbox{\subfigure[\label{bb}]{\includegraphics[width=0.45\textwidth]{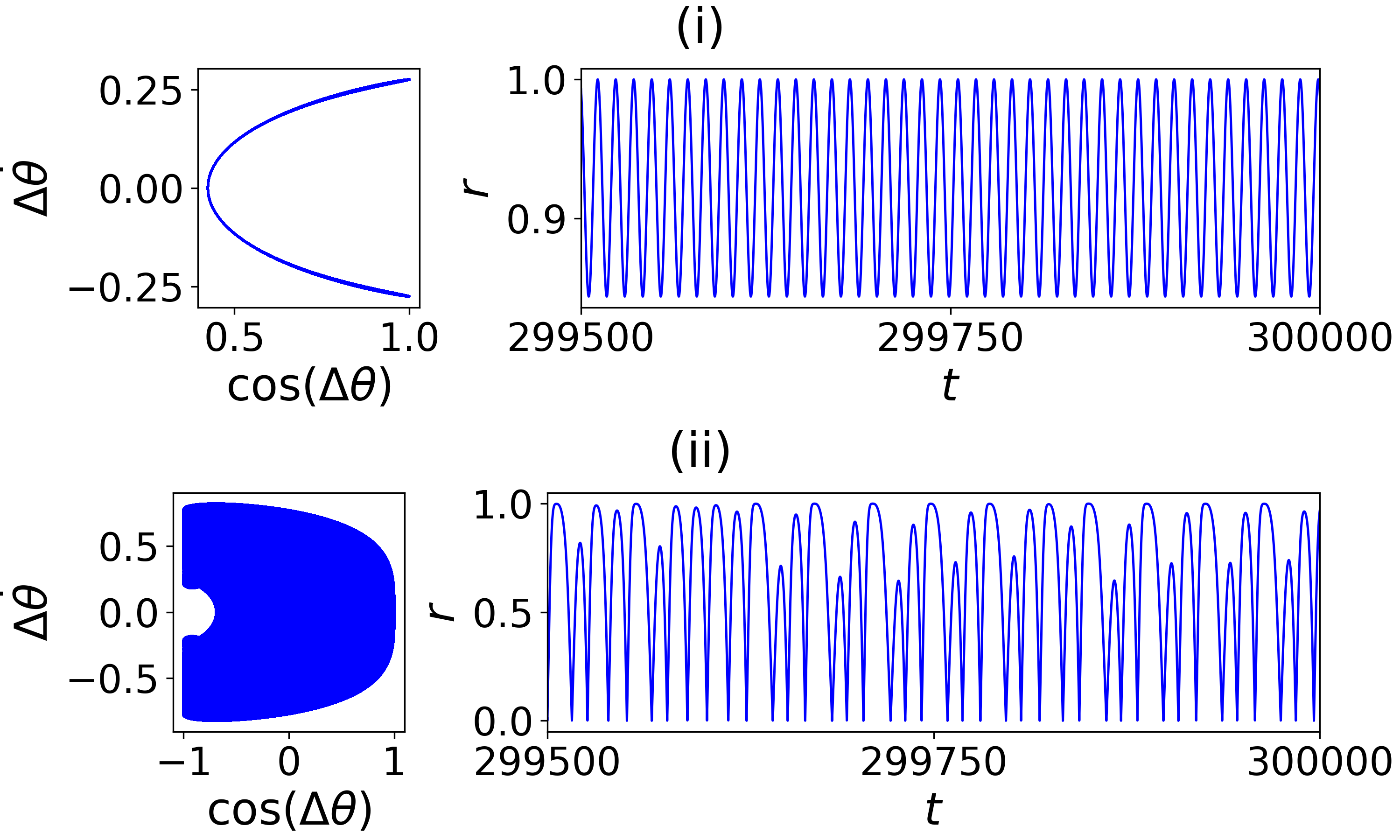}}}
	\fbox{\subfigure[\label{cc}]{\includegraphics[width=0.45\textwidth]{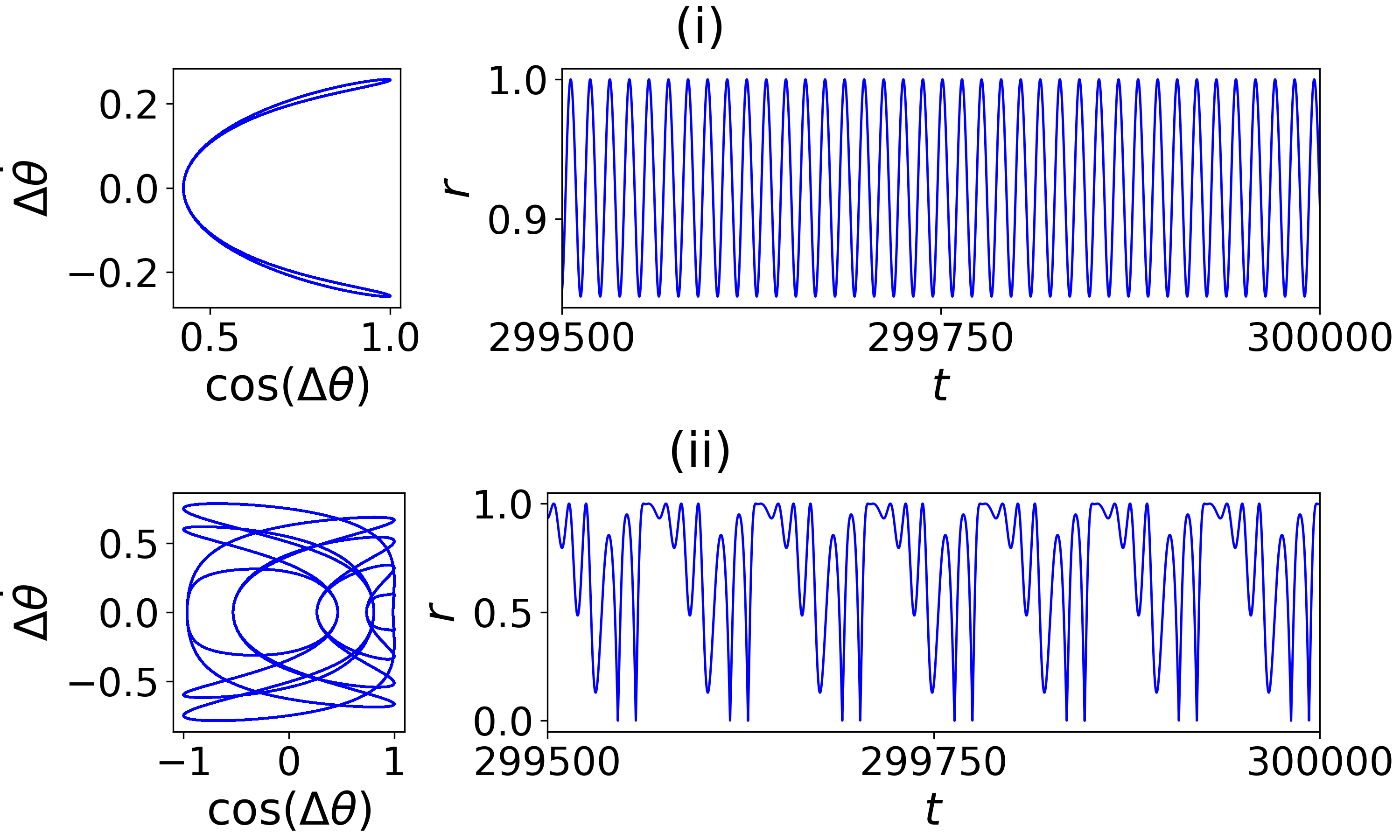}}}
	\fbox{\subfigure[\label{ee}]{\includegraphics[width=0.45\textwidth]{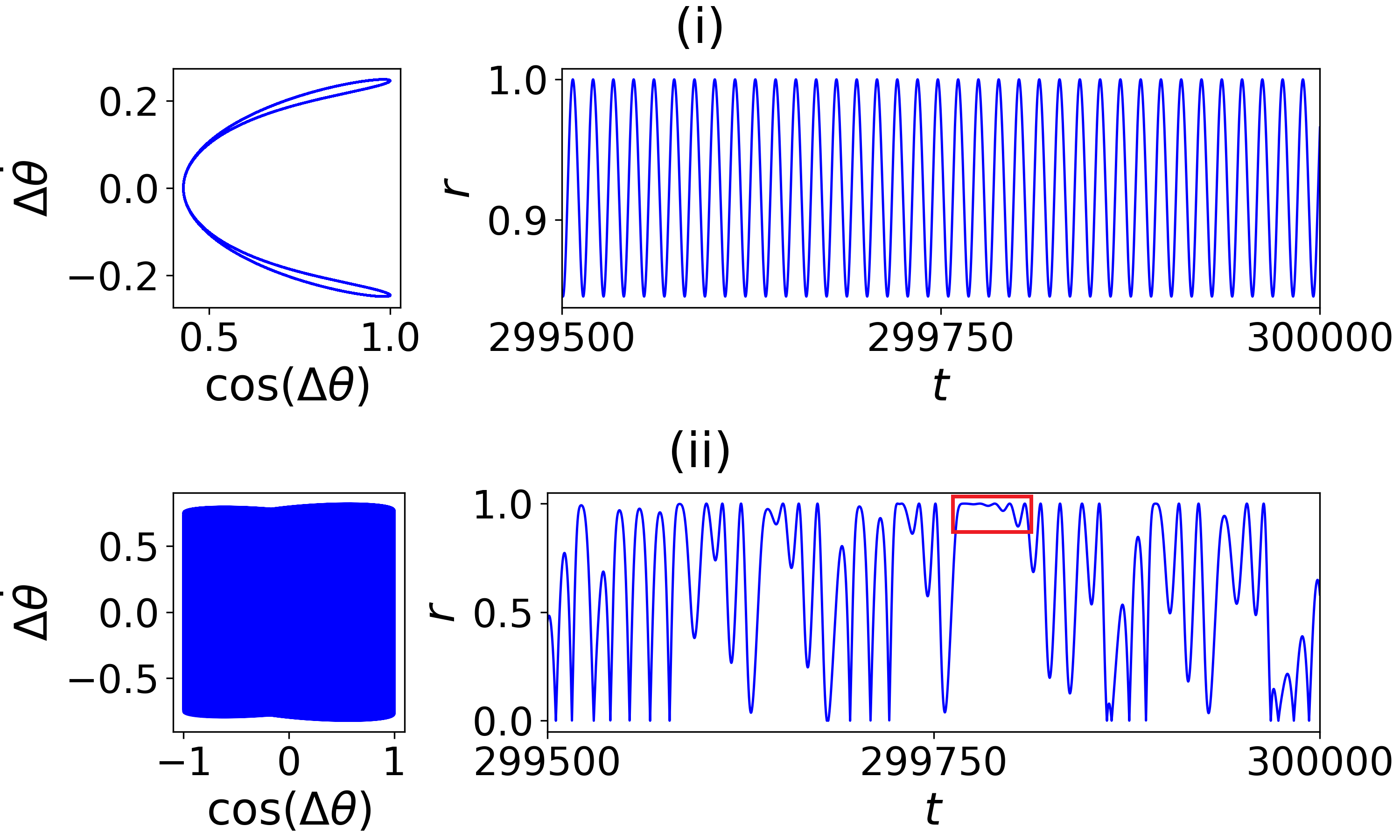}}}
	 \caption{Numerical non-phase-locked solutions for $25$ realizations with $m=10$, $K=1$, and $\omega_1=\omega_2=1$. Different panels correspond to (a) $\tau=3.5$, (b) $\tau=3.8$, (c) $\tau=4.4$, (d) $\tau=4.8$. Each panel shows the phase-space i.e. $\Delta\dot{\theta}={\dot{\theta}}_1-{\dot{\theta}}_2$ versus $\cos(\Delta{\theta})$  for the last $10^6$ time steps (left), and the order parameter $r$ versus time (right). The red box represents patches of intermittent synchrony wherein the system remains nearly fully synchronized just for a specific duration.}
	\label{fig5:numeric2}
\end{figure*}

\begin{figure*}[]
	\centering
	  \subfigure[\label{ccc}]{\includegraphics[width=0.68\columnwidth]{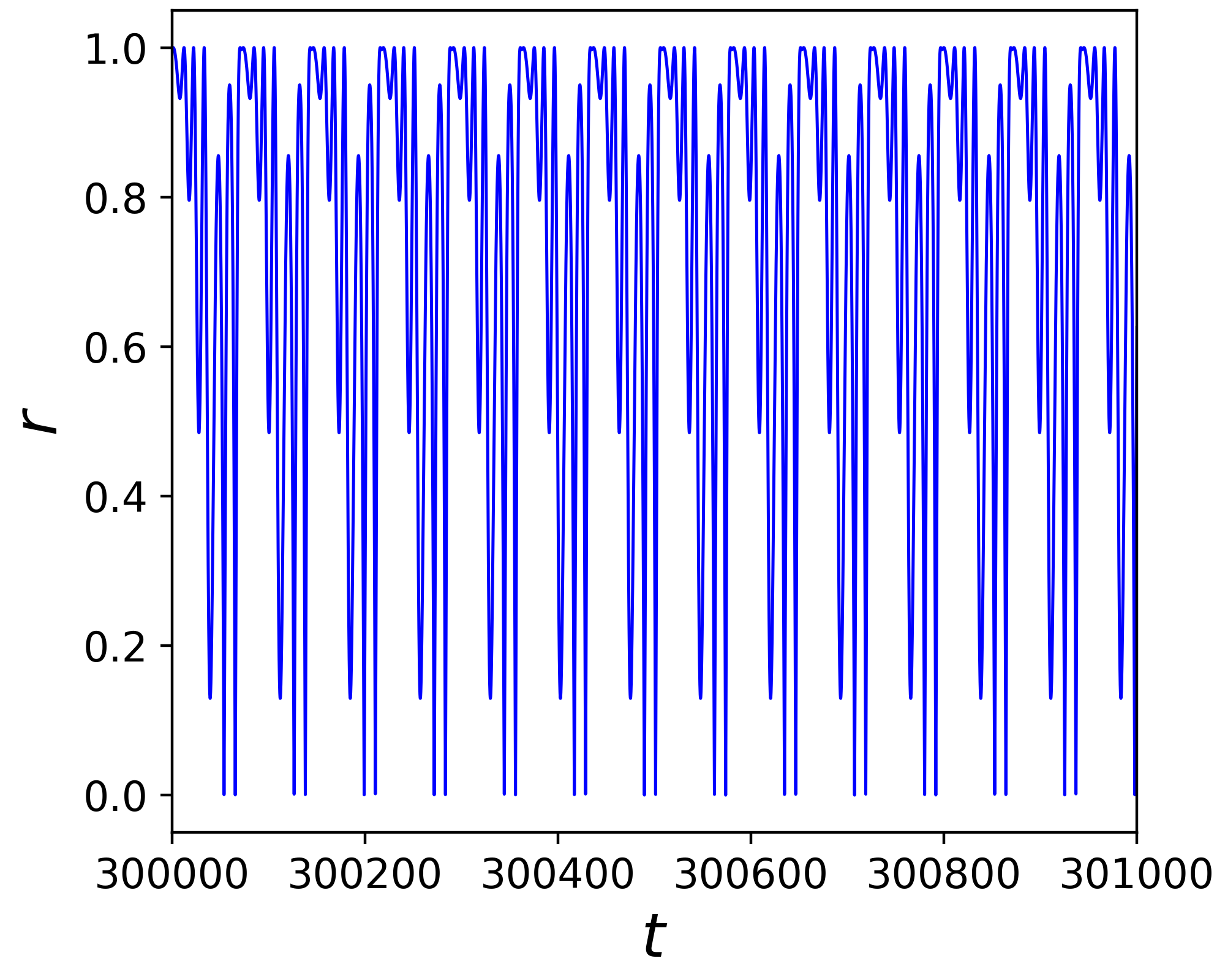}}
	 \subfigure[\label{ddd}]{\includegraphics[width=0.68\columnwidth]{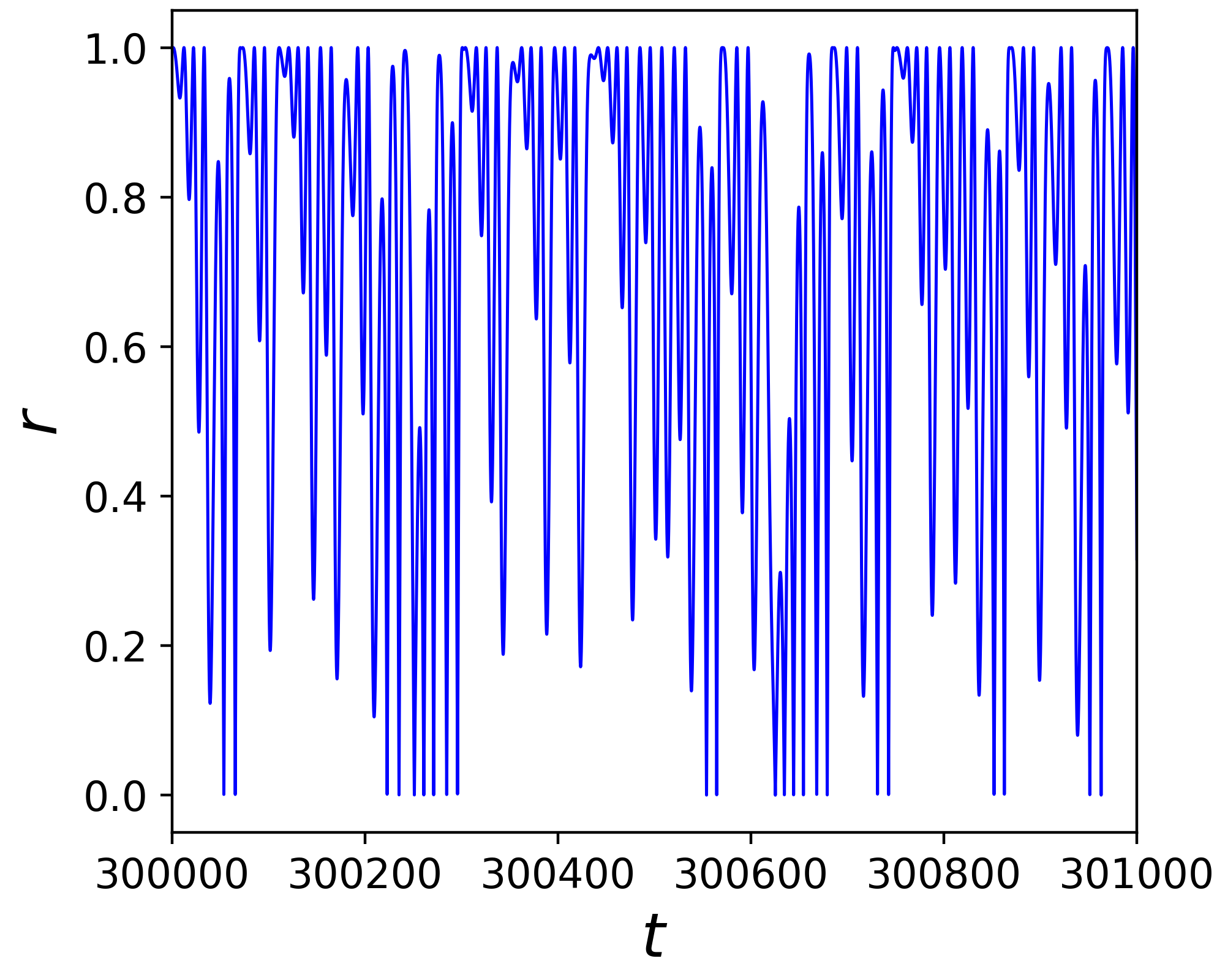}}
	 \subfigure[\label{eee}]{\includegraphics[width=0.68\columnwidth]{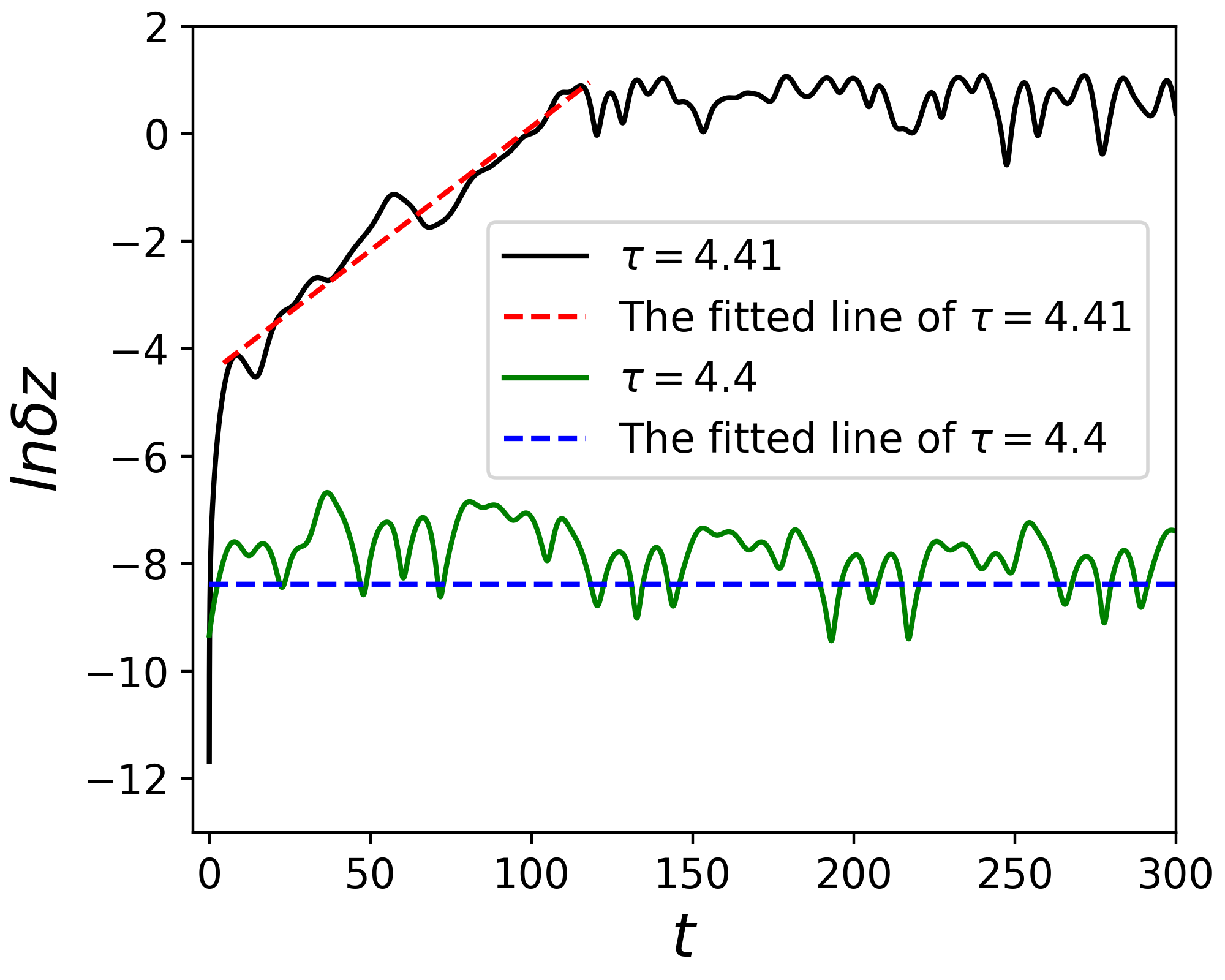}}
	\caption{Temporal variation of the order parameter for two identical phase oscillators with $m=10$, $K=1$, and $\omega_1=\omega_2=1$, and for time delays (a) $\tau=4.4$ and (b) $\tau=4.41$. Panel (c) shows the natural logarithm of trajectory difference $\delta{z}$ versus time for $\tau=4.4$ (green line) and $\tau=4.41$ (black line). The Largest Lyapunov Exponents ($LLE$) are obtained by computing the slope of the linear fit to the early stages of these curves.}
	\label{fig6:parametric-chaos}
\end{figure*}

\begin{figure*}[]
	\centering
  \subfigure[\label{a}]{\includegraphics[width=0.9\columnwidth]{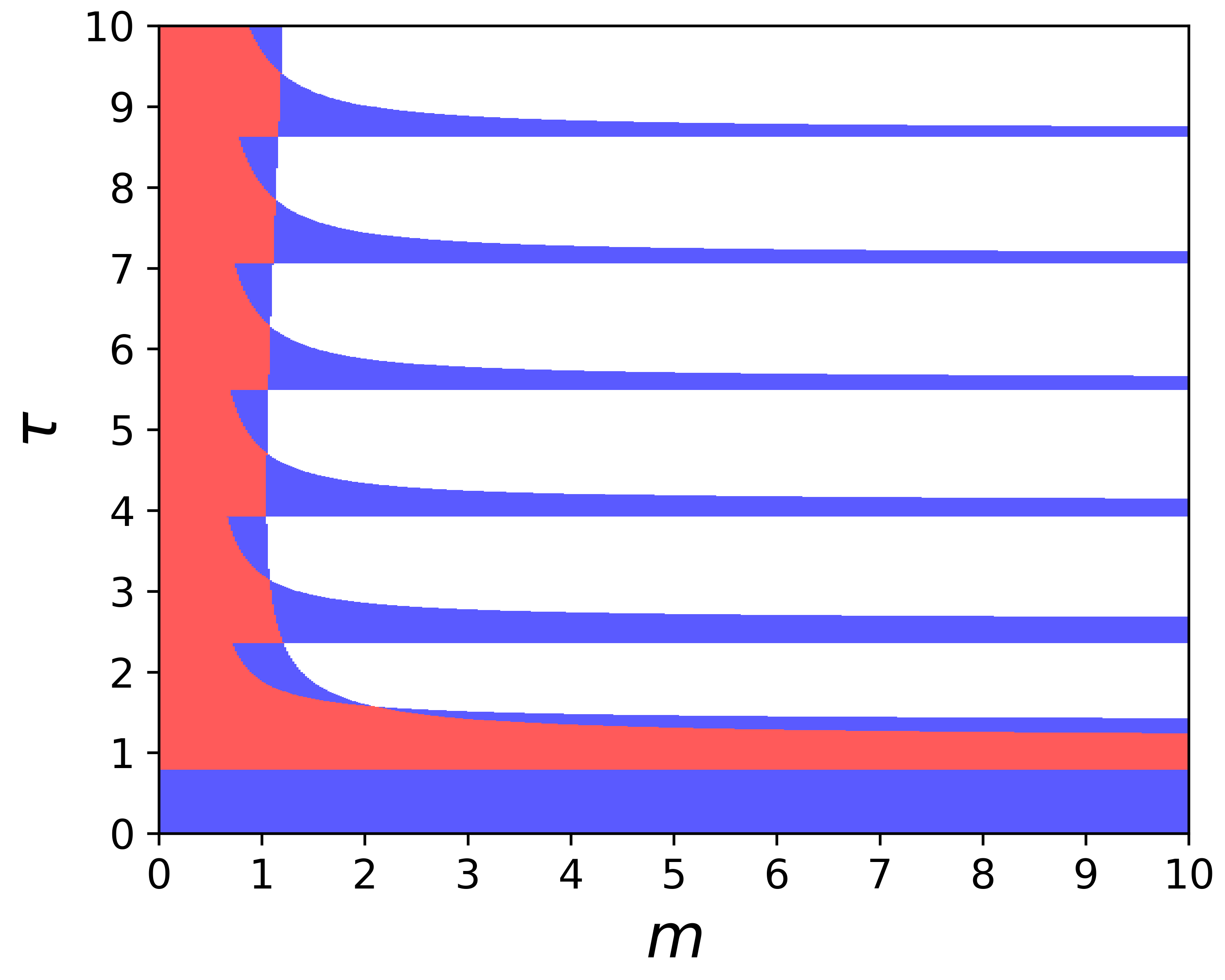}}
	 \subfigure[\label{b}]{\includegraphics[width=0.9\columnwidth]{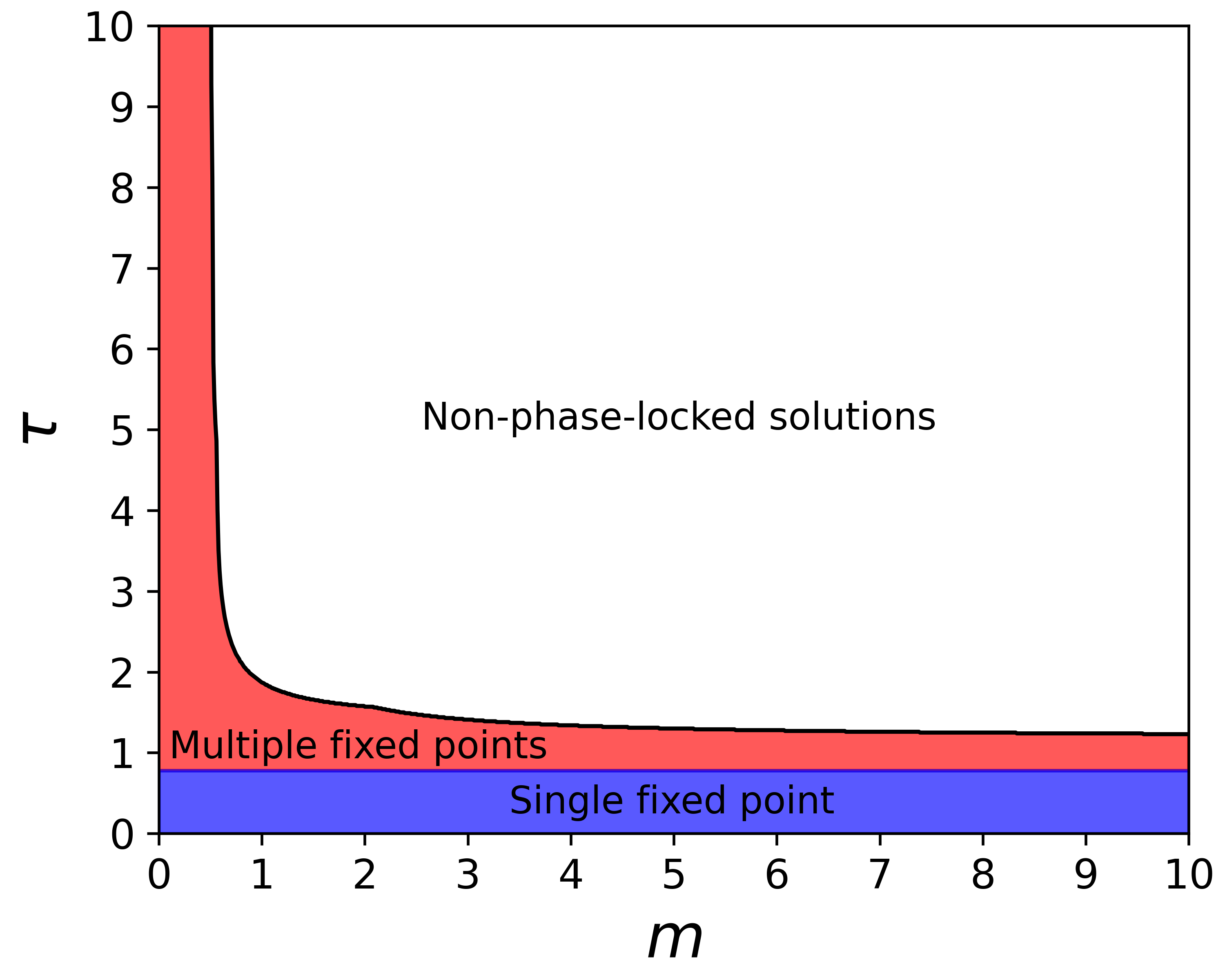}}
	\caption{ (a) analytical and (b) numerical phase diagrams of two identical phase oscillators ($\omega_1=\omega_2=1$). In the diagrams, the colors blue, red, and white represent the regions of single fixed points, multiple fixed points, and non-phase locked solutions, respectively. The numerical phase diagram is obtained through 100 realizations, and $K=1$ is considered for the calculations.}
	\label{fig7:phasediagram1}
\end{figure*}

\begin{figure*}[]
	\centering
	 \subfigure[\label{a}]{\includegraphics[width=0.49\columnwidth]{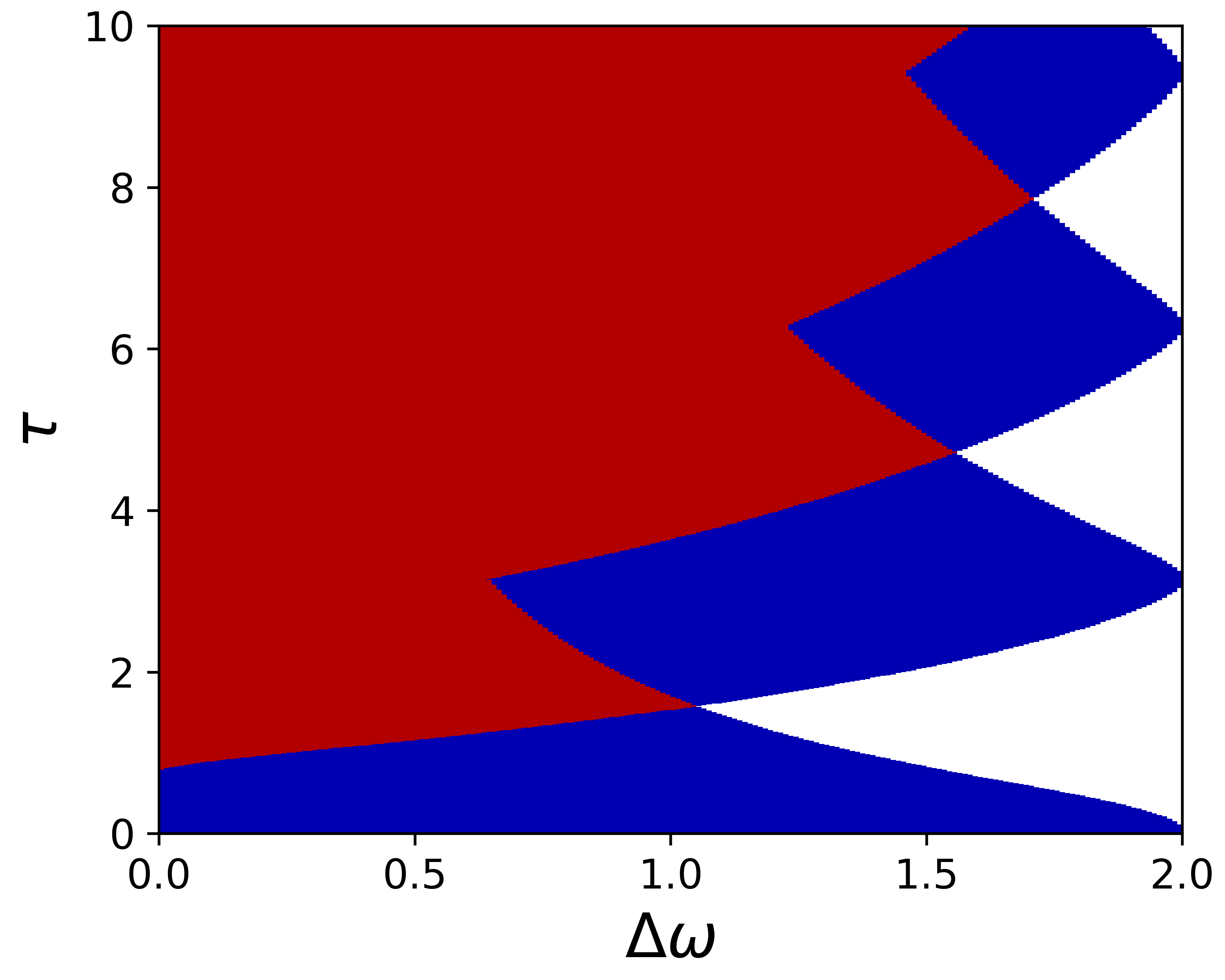}}
	 \subfigure[\label{b}]{\includegraphics[width=0.49\columnwidth]{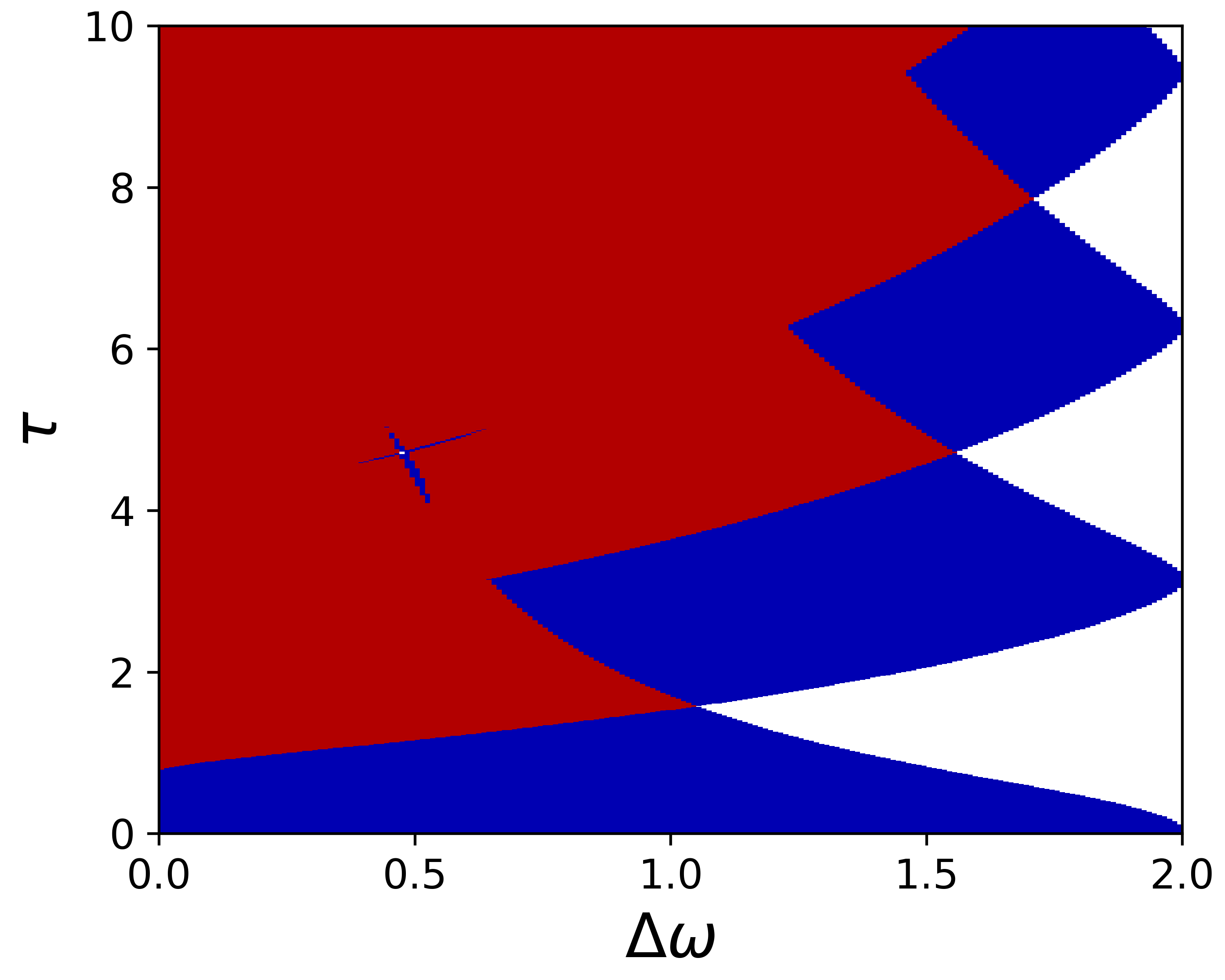}}
	 \subfigure[\label{c}]{\includegraphics[width=0.49\columnwidth]{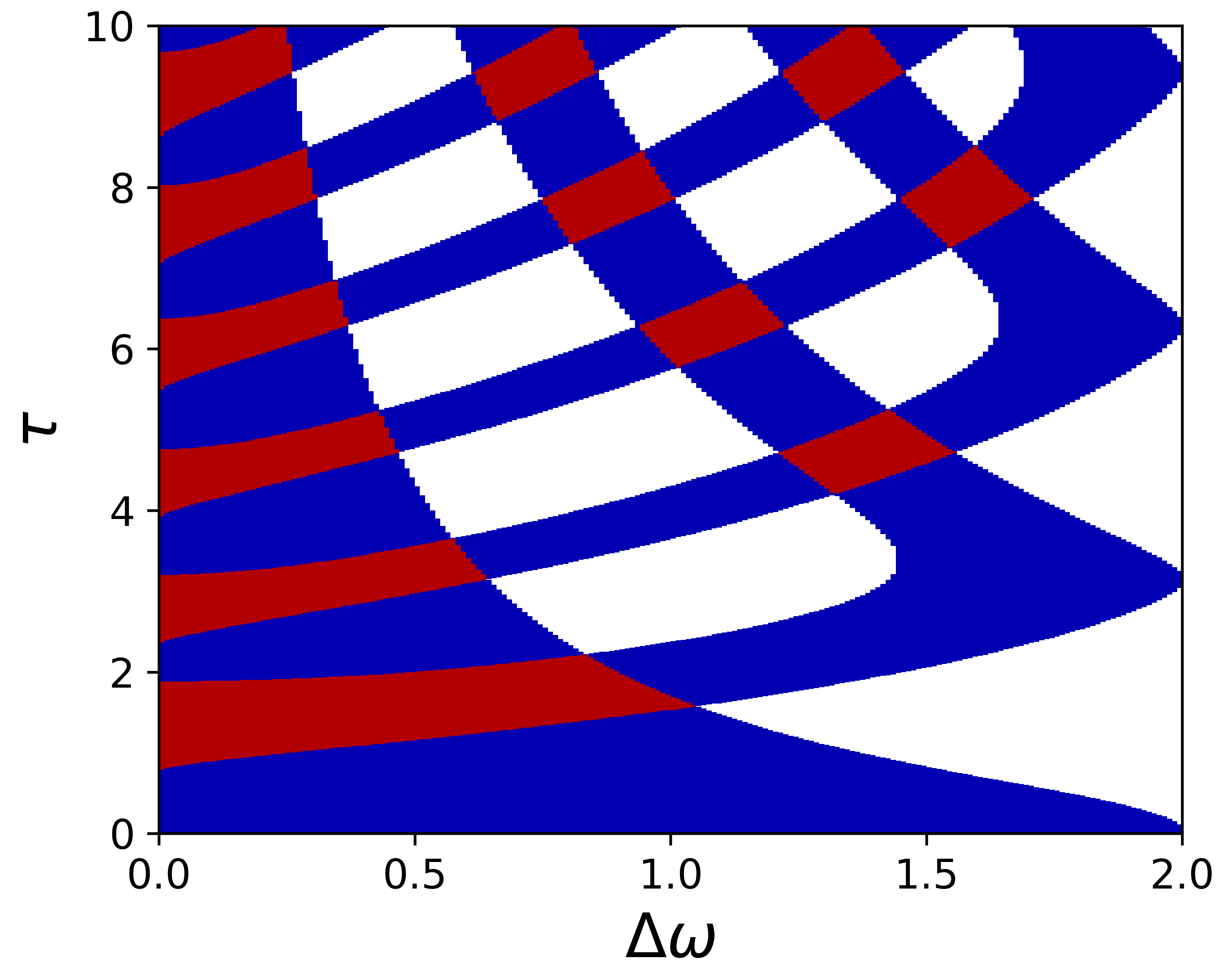}}
	 \subfigure[\label{d}]{\includegraphics[width=0.49\columnwidth]{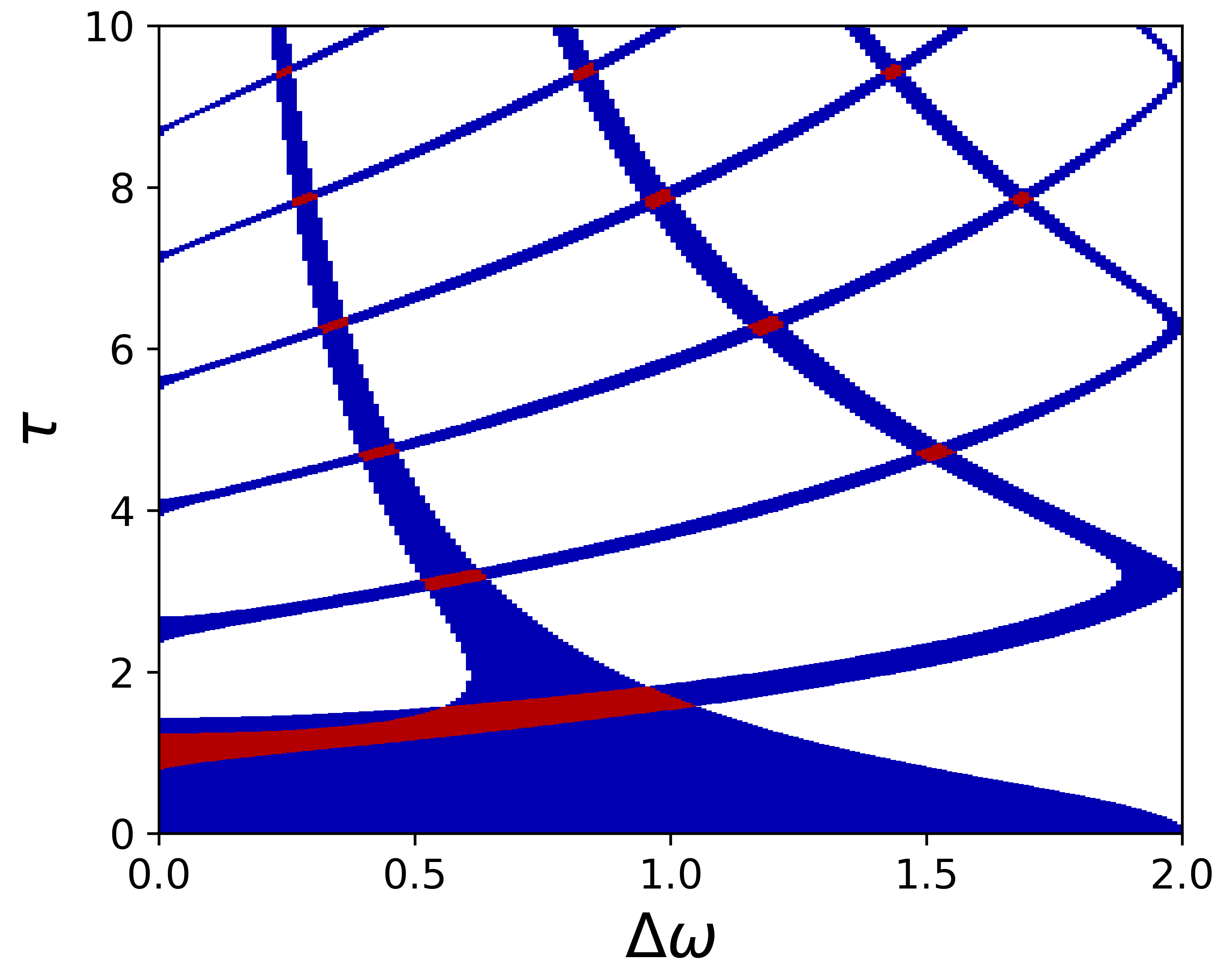}}	
	\caption{ Analytical phase diagram of two non-identical phase oscillators presented in the $\tau-\Delta\omega$ plane for (a) $m=0$, (b) $m=0.6$, (c) $m=1.0$, and (d) $m=10$. The mean frequency is set to unity and $K=1$. The areas colored blue, red, and white correspond to the occurrence of single fixed points, multiple fixed points, and no fixed points, respectively. }
	\label{fig8:phasediagram2}
\end{figure*}
\begin{figure}[]
	\centering
	 \subfigure[\label{a}]{\includegraphics[width=0.49\columnwidth]{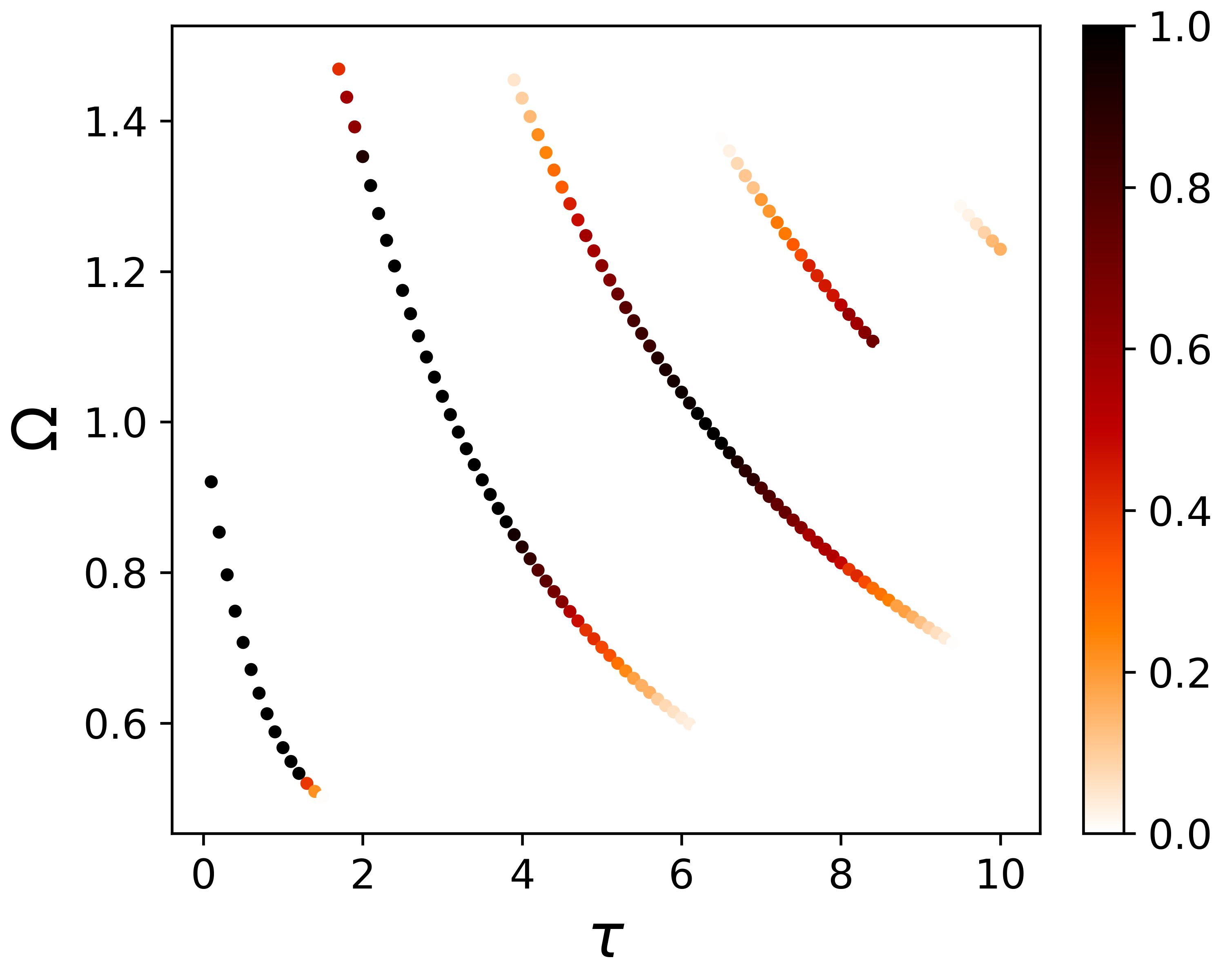}}	
	 \subfigure[\label{b}]{\includegraphics[width=0.49\columnwidth]{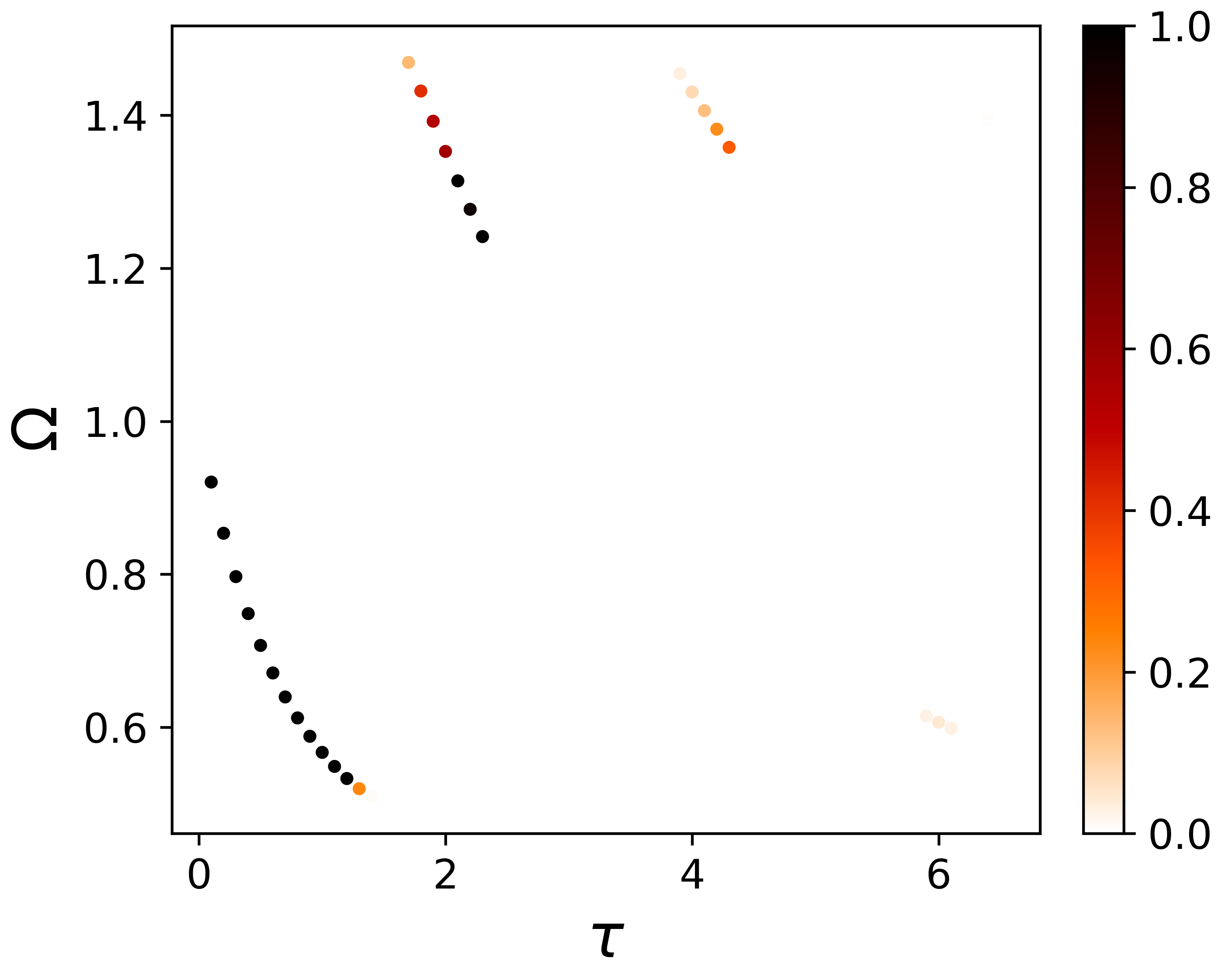}}
	\caption{Probability of phase-locked solutions estimated over 500 realizations with $K=1$, $\omega_1=1.5$, and $
	\omega_2=0.5$, and for (a) $m=0.6$, and (b) $m=1.0$. The probability of each solution is encoded by a color with a value 
	indicated in the sidebar.}
	\label{fig9:numeric}
\end{figure}
\begin{figure}[]
	\centering
	 \subfigure[\label{a}]{\includegraphics[width=0.49\columnwidth]{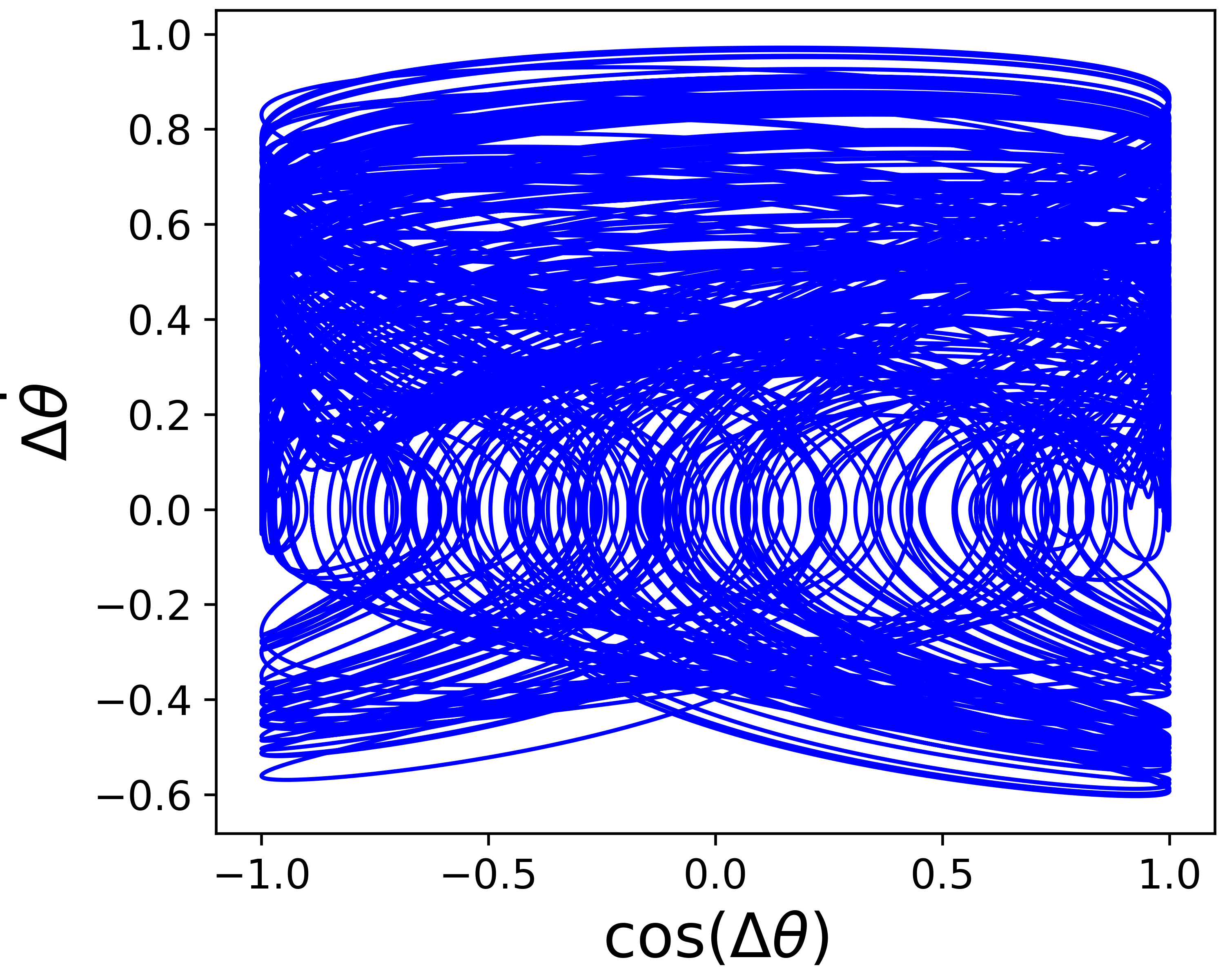}}	
	 \subfigure[\label{a}]{\includegraphics[width=0.49\columnwidth]{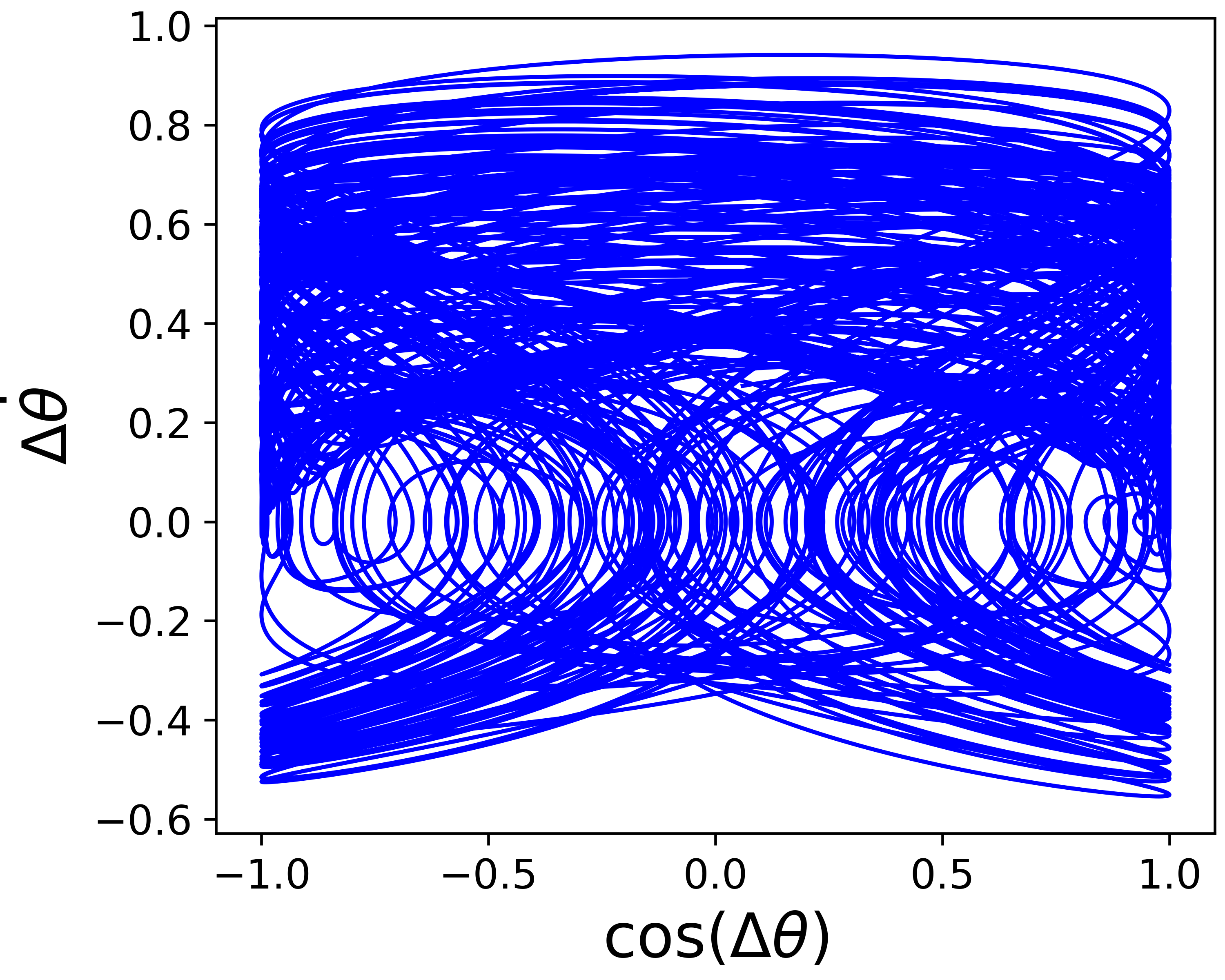}}	
	 \subfigure[\label{a}]{\includegraphics[width=0.49\columnwidth]{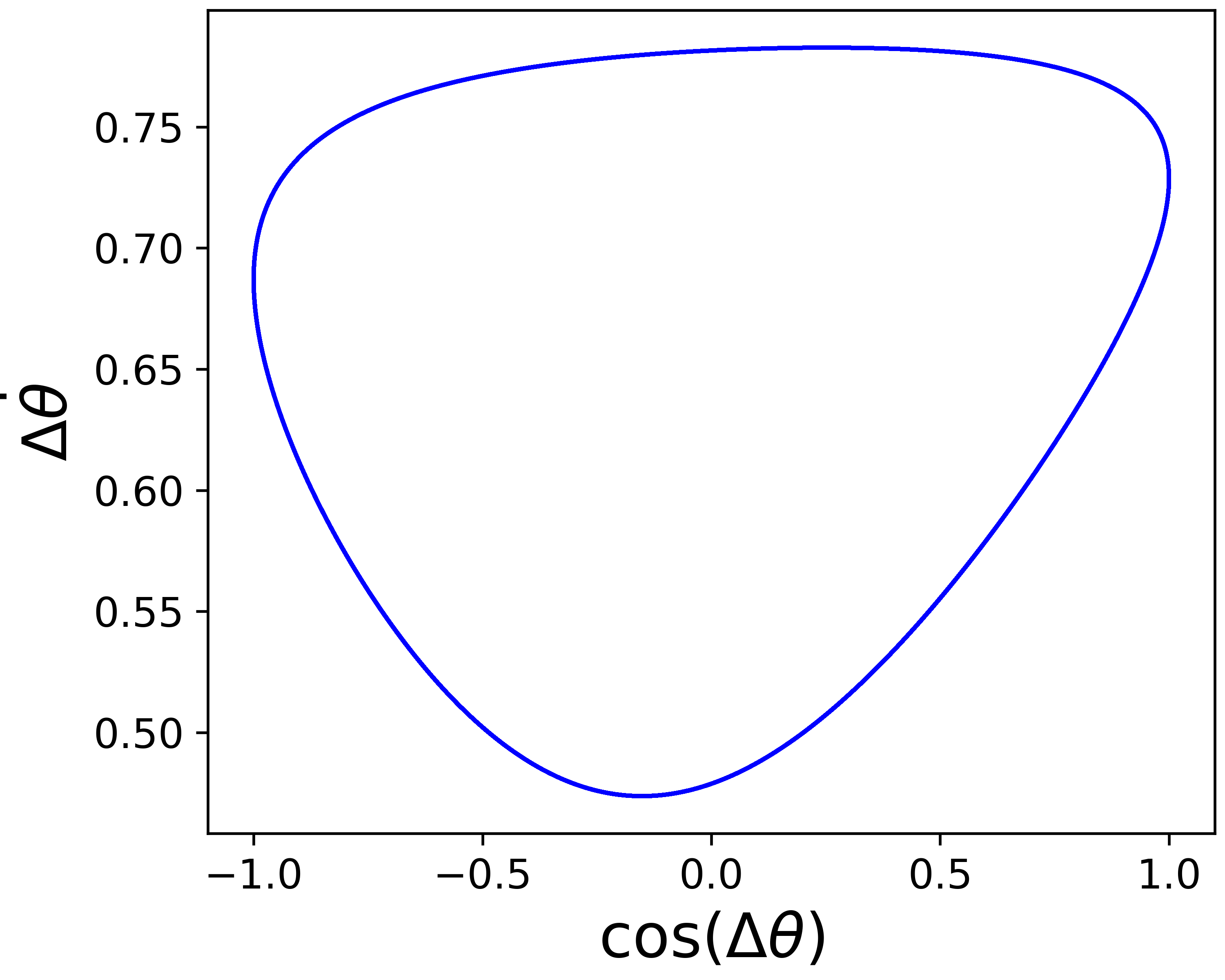}}	
	 \subfigure[\label{a}]{\includegraphics[width=0.49\columnwidth]{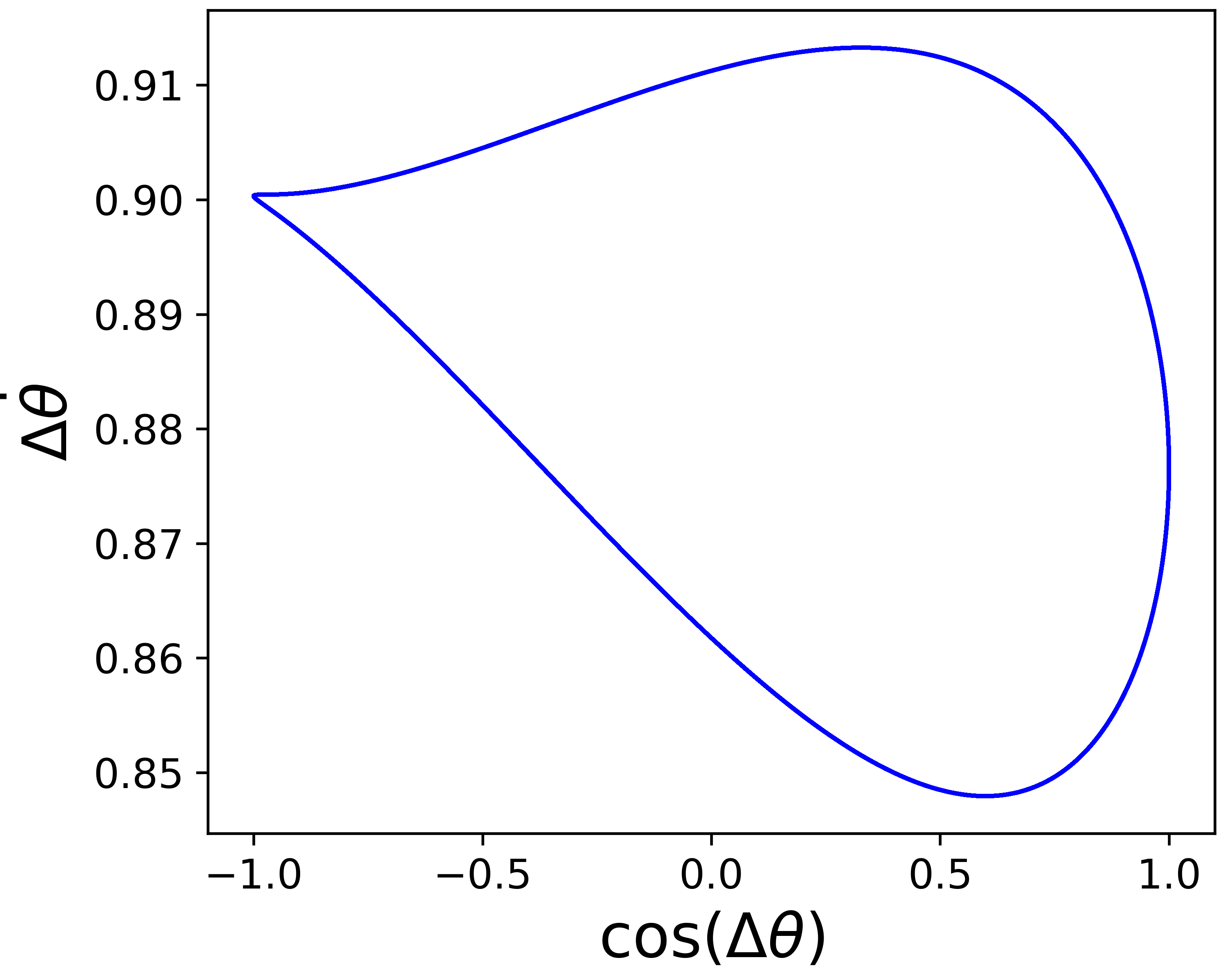}}	
	\caption{Phase-space of two coupled oscillators for $3\times 10^5$ simulations time steps for the frequency differences 
	(a) $\Delta\omega=0.5$,  (b) $\Delta\omega=0.6$, (c) $\Delta\omega=0.7$, (d) $\Delta\omega=0.8$.  The time delay is $\tau=4.8$, $m=10$, $K=1$, and the mean frequency $\bar\omega=1$. The initial phase distribution is the same for all four 
	simulations. }
	\label{fig10:phasespace}
\end{figure}

We will now proceed to investigate the non-phase-locked solutions. We limit our analysis to the case where $m=10$, $K=1$, and the natural frequencies of the oscillators are $\omega_1=\omega_2=1$. The results are shown in Fig.~\ref{fig5:numeric2}, providing insight into the system's behavior under specific parameter values, including various time delays. 
 
 This figure visually represents various dynamical behaviors observed in $25$ different realizations of the system for the time delays $\tau=3.5, 3.8, 4.4, 4.8 $. Each panel contains two plots, the temporal variation of the order parameter $r$, and the phase space which is the relative angular velocity $\Delta\dot{\theta}={\dot{\theta}}_1-{\dot{\theta}}_2$ versus $\cos(\Delta{\theta})$ over the last $10^6$ time steps, utilizing a time step of $dt = 0.01$. This figure indicates the cyclic or chaotic nature of the non-phase-locked solutions.

Figs.~\ref{fig5:numeric2}-(a) and (c) show two distinct periodic behaviors for $\tau=3.5$, and $\tau=4.4$ exhibiting closed paths in phase space and periodic variations in the order parameter. Figs.~\ref{fig5:numeric2}-(b) and (d), however, illustrate the presence of chaotic solutions in addition to the periodic ones for $\tau=3.8$ and $\tau=4.8$, respectively, in which the phase spaces for chaotic solutions are almost filled. It is worth noting that in the chaotic solutions for $\tau=4.8$, the two oscillators tend to become almost fully synchronized at random intervals, which resembles the epileptic seizure in the brain. Even a slight change in the time delay can have a profound impact.

To rigorously evaluate the stability of a limit cycle in response to minor variations in initial conditions, we introduce a small disturbance around the limit cycle and measure the temporal deviation between the perturbed and unperturbed trajectories, expressed as:

\begin{equation}
	\delta z=\sqrt{4\sin^2(\Delta\theta_1/2)+4\sin^2(\Delta\theta_2/2)+{\Delta{\dot{\theta}_1}}^2+{\Delta{\dot{\theta}_2}}^2}.
	\label{Eq:dltz}
\end{equation} 
Here $\Delta\theta_1$ and $\Delta{\dot{\theta}_1}$ ($\Delta\theta_2$ and $\Delta{\dot{\theta}_2}$) denote the phase and frequency differences between perturbed and unperturbed trajectories of the first (second) oscillator. The Largest Lyapunov Exponent ($LLE$) is determined as the slope of variation of $ln\delta{z}$ versus time. 

Figs.~\ref{fig6:parametric-chaos}-(a) and (b) illustrate the evolution of the order parameter under identical initial conditions, for $\tau=4.4$, and $\tau=4.41$, respectively. As observed in the figures, a slight modification in the time delay parameter can lead to a transition from periodic to chaotic behavior in the system. This transition to chaos can be quantified by determining the Largest Lyapunov Exponent. After reaching a steady state at $\tau=4.4$, the delay is adjusted to $4.41$, and the corresponding $LLE$ is determined by examining the slope of the $ln\delta{z}$ vs. time, represented as the dashed-red line in Fig.~\ref{fig6:parametric-chaos} (c). The mean $LLE$ across 20 different perturbations is found to be zero for $\tau=4.4$ and $0.045\pm 0.003$ for 
$\tau=4.41$. The $LLE$ value of zero suggests the stability of the limit cycle for $\tau=4.4$. On the other hand, a positive $LLE$ value indicates the exponential growth of $\delta z$ and therefore the chaotic behavior for $\tau=4.41$. The susceptibility to parameter adjustments emphasizes the complex and nonlinear characteristics of the system's dynamics.


 Figs.~\ref{fig7:phasediagram1}-(a) and (b) display the analytical and numerical phase diagrams of two identical phase oscillators, respectively. Regardless of the value of $m$, the diagram reveals that for $\tau\lesssim 0.79$, there exists only one fixed point attractor. This fixed point corresponds to a phase-locked state, visually represented by the blue regions in the analytical and numerical phase diagrams. 
 However, when $\tau$ exceeds $0.79$, multiple fixed points emerge, forming the red parts. Furthermore, an increase in the time delay leads to the appearance of non-phase-locked solutions (periodic or chaotic) in the white area. While analytical solutions suggest the possibility of single fixed points for large $\tau$ and $m$, they are rarely observed in the 100 realizations used in numerical calculations. As a result, in the numerical phase diagram (Fig.~\ref{fig7:phasediagram1}-(b)), we represent the area beyond the stability regions of single and multiple fixed points with white color. It is worth noting that for $m\lesssim 0.5$, no non-phase-locked solutions are observed across all values of time delay.

\subsection{Case $\omega_1\neq\omega_2$}

 Here, we investigate the phase diagram for the case that the two oscillators have different intrinsic frequencies. Assuming 
$\omega_{1}=1+\frac{\Delta\omega}{2}$ and $\omega_{2}=1-\frac{\Delta\omega}{2}$, 
we set the mean frequency $\bar\omega$ to unity.
Fig.~\ref{fig8:phasediagram2} presents the analytical phase diagram, obtained by the linear stability analysis 
in the $\tau$-$\Delta\omega$ plane. 
This figure illustrates the relationship between the nature of solutions at every time delay and the difference in intrinsic frequencies. The blue, red, and white colors denote the regions with single, multiple, and no fixed points. 
For a small enough time delay, i.e. $\tau<2$, in the interval $0\leqslant\Delta\omega\lesssim 1$, 
one can see an increase in the range of time delays for which only one fixed point exists. However, for $\Delta\omega >1$ 
the region of single fixed point solutions shrinks, and the regions with
 non-phase-locked solutions appear, and their sizes enlarge by increasing $\Delta\omega$. Increasing inertia, i.e. $m\geq 1$, tends to narrow the single and multiple fixed point regions and the manifestation of non-phase-locked solutions even for small values of $\Delta\omega$. 
For a given time delay with a small value of m, as the frequency difference increases, the number of fixed points decreases. Initially, there are multiple fixed points, which decrease to a single fixed point, and eventually none. However, for large values of m, increasing $\Delta\omega$ may restore the stability of some fixed points.
 
To check the effect of frequency difference on the stability of fixed points, we investigated the numerical solutions over 500 realizations for $m=0.6, 1$ and $\Delta\omega=1$. The results, illustrated in Figs.~\ref{fig9:numeric}-(a) and (b), indicate that the probability of observing the fixed points extends to the higher values of time delays comparing the case of identical oscillators displayed in 
Figs.~\ref{fig4:numeric1}-(b) and (c). 

Finally, We examine how differences in frequency affect the characteristics of non-phase-locked solutions. Our findings show that starting from the same phase distributions, increasing $\Delta\omega$ can change a chaotic solution into a limited-cycle one. For example, Figs.~\ref{fig10:phasespace}-(a) to (d) illustrate the phase-space of two oscillators with $m=10$ and $\tau=4.8$, and  $\Delta\omega=0.5, 0.6, 0.7, 0.8$, respectively. This figure demonstrates the transition from chaotic to periodic behavior with a small change in the frequency difference. We repeated this simulation for 25 different initial conditions and observed such chaotic to periodic transitions in all realizations.


\section{conclusion}
\label{conclusion}

In summary, we investigated how time delay affects the second-order Kuramoto model for two coupled oscillators and its relevance to neurological disorders. When considering two identical phase oscillators, our results show that increasing the time delay leads to the appearance of multiple stable phase-locked solutions. Additionally, we observed that higher inertia coefficients result in fewer stable phase-locked solutions. Furthermore, our research revealed that as the difference between the frequency of each stable state and the average frequency ($\bar\omega$) increases, the size of its basin of attraction decreases.

Our study also illustrated that the presence of time-delayed interaction in the second-order Kuramoto model can generate non-phase-locked solutions, which can lead to limit-cycle or chaotic behaviors based on the inertia coefficient and the amount of time delay. Our findings suggested that the observed behavior is highly sensitive to the value of the time delay. Even a slight change in the time delay can cause a shift from periodic to chaotic behavior. Our findings also show that the difference in interinsic frequencies expands the stability range of fixed point solutions to longer time delay values. Additionally, in the case of non-fixed point solutions, the increase in frequency disparity can cause a shift from chaotic to regular dynamics.

For systems with significant inertia and time delay, behaviors resembling epileptic seizures can be observed in the system's dynamics. These behaviors occur when the oscillators are nearly fully synchronized for a period of time before eventually going out of phase. A previous study by Gerster et al.~\cite{gerster2020fitzhugh} reported the spontaneous occurrence of synchronization phenomena that closely resemble the observed behavior during epileptic seizures. This occurrence is attributed to the complex interaction between dynamics and network topology. The impact of time delay is significant in the context of epilepsy, as seizures are often observed as a symptom in demyelinating diseases, such as multiple sclerosis~\cite{de2021hypothesis}. Indeed, the transmission of information in neuronal systems is influenced by several factors, one of which is the presence of myelin. Myelin is a fatty substance that encircles axons, providing an insulating layer that boosts the speed and efficiency of signal propagation. If myelin is lost or damaged, such as in certain neurological disorders or injuries, the transmission of signals can slow down significantly, leading to a longer time delay between the interactions of neurons. These observations suggest a potential connection between the dynamics of time-delayed coupled massive oscillators and the pathological synchronization observed in epileptic seizures and demyelinating diseases.

It has been shown that the Josephson junctions can be used to model the dynamics of a single neuron~\cite{mishra2021neuron}.  
A simplified model of a neuron with an axon and dendrites can be represented as a two-dimensional phase oscillator with inertia. This model is capable of explaining the intrinsic transient behavior commonly observed in experimental settings of the networks of dendritic neurons~\cite{dolan2005phase,majtanik2006desynchronization,sakyte2011self}. Our results also indicate that the second-order Kuramoto model may be more suitable for describing the behavior of neurons that the original Kuramoto model cannot capture. Further research is necessary to explore this connection.

Overall, the study of coupled time-delayed massive oscillators provides a valuable framework for understanding the complex behaviors observed in many natural and engineered systems. Therefore, we hope that our findings will contribute to a deeper understanding of the underlying mechanisms that govern their behavior.

\section*{Data Availability Statement}
Data sharing is not applicable to this article as no datasets were generated or analyzed during the current study.

\section*{References}
\bibliography{bibliography2}

\end{document}